# On the probability of down-crossing and up-crossing rogue waves

Alexey V. Slunyaev[1-3)] and Anna V. Kokorina[2)]

[1)] National Research University Higher School of Economics, Nizhny Novgorod, Russia
[2)] Institute of Applied Physics of the Russian Academy of Sciences, Nizhny Novgorod, Russia
[3)] V.I. Il'ichev Pacific Oceanological Institute, Far Eastern Branch RAS, Vladivostok, Russia

**Abstract**
By means of the direct numerical simulation of directional waves on the surface of deep water it is shown that extreme waves can exhibit such asymmetry that the occurrence of deeper troughs is several times more likely on the wave rear slopes. This effect becomes most pronounced in the case of steep short-crested waves. It is not related to the Benjamin – Feir instability, but is a result of complex contribution from nonlinear combination harmonics, mainly cubic in nonlinearity. The discovered asymmetry can lead to remarkably different estimates of the rogue wave probability based on either down- or up-zero-crossing methods for individual wave selection, commonly used in the oceanography.

## 1. Introduction

In this work we discuss a specific geometric feature of water waves from the population of most extreme irregular waves in a deep sea. This issue is related to the problem of identification of the characteristic portrait of so-called rogue waves in the ocean, which are commonly defined as abnormally high waves that occur suddenly in stochastic fields of wind-generated waves. Though oceanic rogue waves are certainly rare events, they represent a real threat, which is further exacerbated thanks to the rapid development of the sea use for various human needs, such as transportation, off-shore installations, recreation, etc. A huge number of recent scientific studies are dedicated to various aspects of the rogue wave problem, see the reviews [Dysthe et al., 2008; Kharif et al, 2009; Onorato et al., 2013b; Slunyaev et al., 2023].

Return periods of waves of a given height may be described by the probability distribution function (PDF). The classic theory for a system of linear waves with random phases and narrow spectrum is represented by the Gaussian statistics for the water surface displacements, and the Rayleigh distribution for the wave height exceedance probability [Massel, 1996; Holthuijsen, 2007]

$$P(H) = \exp\left(-2\frac{H^2}{H_s^2}\right), \qquad (1)$$

where the wave height $H$ is normalized by the representative, so-called significant, wave height $H_s$. Surprisingly, this simple formula, which depends solely on the wave height ratio, is able to accurately describe oceanic waves in typical situations. The rogue wave population is most frequently specified using the criterion on the amplification index $AI = H/H_s$, requiring $AI > 2$.

Similar to (1) distributions describe the probabilities of wave crest and wave trough amplitudes with respect to the mean water level. The complex interplay between nonlinear and dispersive effects makes real sea waves different from the idealized representation in the form of independent sinusoidal random waves suggested by Longuet-Higgins (1952). The distribution (1) is affected by nonlinearity, and this effect is potentially stronger for rare waves with large amplification indices *AI*.

The wave nonlinearity influences most substantially the distributions of wave crest and wave trough amplitudes, $A_{cr}$ and $A_{tr}$ respectively; first of all – due to bound waves which modify the individual wave shape. The second-order nonlinear correction to the sinusoidal wave shape is produced by the phase-locked (also known as bound) second wave harmonic. It is taken into account in generalizations of the distribution (1), like the popular Foristall (1978) or Tayfun (1980) distributions, and other second order statistical theories. The height of an individual wave may be defined as the sum of the crest and trough amplitudes, $H = A_{cr} + A_{tr}$, where $A_{cr} > 0$ and $A_{tr} > 0$ (see Fig. 1). Then it is easy to see that in the case of a monochromatic wave, the second harmonic does not affect the wave height value, as its contributions to $A_{cr}$ and $A_{tr}$ cancel out. Hence, the wave height is expected to depend on the wave nonlinearity to a lesser extent than the amplitudes of the wave crest and the wave trough, when the wave spectrum is relatively narrow.

The wave-structure interaction features are largely determined not only by the individual wave geometry, but also by the form of the wave sequence, which is influenced by the wave nonlinearity too. The nonlinear self-modulation effect of waves in deep water due to the Benjamin – Feir instability leads to the generation of specific wave patterns known as envelope solitons and breathers, which were introduced to the sea-keeping test practice in [Onorato et al., 2013a; Klein et al., 2021]. Thus, the realistic picture of extreme waves and their group structure are of practical importance.

The typical rogue wave shape has been already discussed in theoretical and experimental works. The researchers are unanimous that extremely high waves in deep water condition are most frequently represented by a high narrow crest surrounded by shallower troughs, what is consistent with the classic picture of the nonlinear Stokes wave, for which the crest-to-trough asymmetry increases with wave steepness. The Stokes wave solution is known to be fully symmetric with respect to the vertical line until the limit when it becomes unstable and overturns.

From the analysis of 1.1 billion individual waves with almost 75 thousand rogue waves satisfying the criterion *AI* > 2, presented by Cattrell et al. (2018), one may conclude that rogue waves typically have the front trough deeper than the rear trough, see their figure 6. Such a wave shape is represented in our Fig. 1a. A similar waveform was suggested in [Lavrenov, 1998] as a particular dangerous feature of rogue waves occurring near the south-western coast of Africa against the strong Agulhas current. At the same time, the results of measurements near the Brazilian coast [Pinho et al., 2004] revealed the opposite picture, that rogue waves with the deeper trough behind the extreme crest were encountered almost twice as often as those with the deeper troughs in front; the dominating rogue wave shapes were similar to the ones shown in our Fig. 1b. Meanwhile, another study of an impressive amount of data consisting of 122 million individual waves [Christou & Ewans, 2014] did not report any noticeable asymmetry between troughs from the face and back sides of the rogue wave crests.

The difference between depths of the preceding and following troughs is directly related to the difference in the wave height PDFs obtained according to the down-zero-crossing and up-zero-crossing methods, which are two standard approaches in the modern oceanographic practice [Massel, 1996; Holthuijsen, 2007], see Fig. 1. In particular, [Pinho et al., 2004] concluded that the number of up-crossing rogue waves in the instrumental time series (see the left part of Fig. 1b) was remarkably greater than the number of down-crossing rogue waves (see the left part of Fig. 1a). So far, in-situ measurements seem to be unable to give a definite answer on the probability of extreme sea waves of the rogue wave type. Therefore, the physical modeling in laboratory experiments and the direct numerical simulation within the primitive equations of hydrodynamics play very important roles in this research.

The average unidirectional rogue wave profile was depicted in [Xiao et al., 2013] on the basis of extensive direct numerical simulation of irregular sea waves. The simulation was performed by means of the High Order Spectral Method (HOSM) applied to the Euler equations for potential flows. The obtained typical rogue wave shape was compared with the symmetric waveform in the vicinity of the most extreme crest within the linear theory for Gaussian waves [Lindgren, 1970; Boccotti, 1983]. Xiao et al. (2013) concluded that in the nonlinear numerical simulations the rogue wave troughs were shallower from the front side and deeper from the back side, in qualitative agreement with the measurements by Pinho et al. (2004), see Fig. 1b. The relation between wave group shapes (in particular, the group set-up or set-down) and the wave directionality was discussed in [Adcock & Taylor, 2009; Adcock, 2017; Slunyaev et al., 2017].

In our own previous stochastic numerical simulations of unidirectional and directional waves with the JONSWAP frequency spectrum in the deep water conditions [Sergeeva & Slunyaev, 2013; Slunyaev & Kokorina, 2020b], and also unidirectional waves in intermediate depths [Slunyaev et al., 2016] using the HOSM code, we found that rogue waves with deeper troughs following the extreme wave crests were dominating in number (similar to Pinho et al. (2004) and Xiao et al. (2013)), when the average wave steepness was sufficiently high. (Note that the schematic wave shapes in figure 5 of [Slunyaev et al., 2016] were swapped by a mistake, unfortunately.) Such rogue wave asymmetry was not observed in the irregular wave states characterized by weaker nonlinearity.

The Benjamin – Feir instability of deep-water waves with respect to long modulations is a threshold nonlinear effect, which requires the average wave steepness to be of sufficiently large magnitude [Johnson, 1997; Kharif et al., 2009]. This modulational instability effect is believed to be a regular mechanism of rogue wave generation under certain spectral conditions [Onorato et al., 2009, 2013; Slunyaev et al., 2023]. As a consequence of the effect, sufficiently long groups of intense waves spontaneously form shorter packets of even steeper waves. Meanwhile, intense wave groups become skewed due to the effect of nonlinear dispersion, which is taken into account by, for example, the fourth-order Dysthe theory for modulated waves, see e.g. [Trulsen, 2007]. Therefore, the Benjamin – Feir instability may be reasonably considered among the first candidates for the role of the mechanism producing the described above asymmetry between troughs adjacent to rogue wave crests. At the same time, no asymmetry in the average rogue wave profile along the wave course was found in the numerical simulations of the directional Dysthe equation in [Socquet-Juglard et al., 2005]. In the simulations of the potential Euler equations [Slunyaev & Kokorina, 2020b] we found that the concerned rogue wave asymmetry in a single-mode wave system with the JONSWAP spectrum grows, when the wave directionality increases. This observation does not seem to be consistent with the well-known picture of the Benjamin – Feir instability of water waves, which quickly vanishes as the angular spectrum broadens [Onorato et al., 2009]. Thus, the relation of the Benjamin – Feir effect to the difference in depths of the troughs preceding and following rogue waves can be questioned.

In this work we systematically address the issue of the asymmetry between rogue wave troughs by means of the numerical simulations of directional nonlinear gravity water waves with the JONSWAP spectrum, which freely evolve on the surface of a deep sea. The simulations are performed within the Euler equations for potential wave motions, solved using the High Order Spectral Method [West et al., 1987] with controlled order of nonlinearity. We not only quantify the balance between rogue waves with deeper troughs located either in front or behind the large crests, but also investigate the origin of this type of wave asymmetry.

In our previous works based on numerical simulations [Sergeeva & Slunyaev, 2013; Slunyaev et al., 2016; Slunyaev & Kokorina, 2020b] the rogue wave populations were obtained applying the condition $AI > 2$, and then were divided into four classes of waveforms depending on the asymmetry about the horizontal and vertical axes. In the present research we follow an alternative approach and calculate two versions of the probability distribution functions, following the down-zero-crossing and the up-zero-crossing methods (see e.g., [Massel, 1996; Holthuijsen, 2007]), as shown in Fig. 1. This approach allows us to consider the difference between wave probabilities in a broad range of wave amplifications $AI$, though still focusing on rogue waves. This approach is consistent with the traditional oceanographic routines.

Typically, in-situ surface wave measurements are represented by time series, when the surface displacement with respect to the mean sea surface level is given in the form of a function of a discrete time, $\eta(t)$. Instead, due to some reason explained below, the main part of the present study deals with space series, $\eta(x)$, where the axis $Ox$ is directed along the dominant wave propagation. Note that then, the front of the same wave appears from the left in the time series, and from the right in the space series. Therefore, a down-crossing wave in the time series will be treated as an up-crossing wave in the corresponding space series and vice versa, cf. the left and right parts of sections in Fig. 1. In order to avoid possible confusion, waves with higher forward slopes (deeper front troughs) will be called hereafter *frontal rogue waves* (Fig. 1a), and, correspondingly, waves with higher back slopes (deeper rear troughs) will be referred to as *rear rogue waves* (Fig. 1b).

The structure of the paper is following. In Sec. 2 we briefly represent the approach which is employed to model the dynamics of nonlinear directional oceanic waves in the direct numerical simulation, and to produce the statistical data. The method of wave decomposition into different nonlinear and directional wave components is described in Sec. 3. It allows to synthesize the wave fields which correspond to particular wave harmonics or their combinations. Some details on the interpretation of the spectral domain within the broad-band second-order nonlinear wave theory are given in the Appendix. We build wave height probability distribution functions for the full surface displacement and for composite wave fields in Sec. 4. This way, we estimate the contribution of different wave components to the probability distributions of down-crossing and up-crossing wave heights. The outcomes of the study are further discussed and summarized in the last section.

## 2. Direct numerical simulation of wave ensembles

Ensembles of irregular directional waves on the surface of a deep water basin are simulated using the High Order Spectral Method [West et al., 1987] which computes the wave evolution in time within the standard set of hydrodynamic equations for potential flows of the ideal fluid formulated in terms of the surface displacement and the surface velocity potential as suggested by Zakharov (1968). The nonlinear surface boundary conditions are calculated approximately using the Taylor expansion for the hydrodynamic fields in the vicinity of the rest water level with a predefined order $M$, which is the nonlinear parameter of the code. The numerical scheme with parameter $M$ accurately describes the nonlinear wave interactions between up to $M + 1$ waves; it is conservative, fully dispersive and preserves the Hamiltonian structure [West et al., 1987; Onorato et al., 2007]. The choice of the parameter $M = 3$ corresponds to the simulation up to four-wave nonlinear interactions, and is used in the literature most frequently. The present simulations are performed for two-dimensional surface displacements $\eta(\mathbf{x},t)$, where $\mathbf{x} = (x, y)$ and the coordinate $x$ corresponds to the dominant direction of wave propagation, $y$ is the transverse coordinate, and $t$ is time. The boundary conditions are periodic in both horizontal directions. The advancing in time is performed using the exact analytical formulas for the linear part and using the 4-order Runge-Kutta scheme with a fixed time step for the nonlinear part of the equations, see details in [Slunyaev & Kokorina. 2020a,b].

The initial conditions at $t = 0$ are specified for the target frequency spectrum of the JONSWAP shape

$$S(\omega) \propto \left(\frac{\omega}{\omega_p}\right)^{-5} \exp\left[-\frac{5}{4}\left(\frac{\omega}{\omega_p}\right)^{-4}\right] \gamma^{\exp\left[-\frac{1}{2\delta^2}\left(\frac{\omega-\omega_p}{\omega_p}\right)^2\right]}, \quad \delta(\omega) = \begin{cases} 0.07, & \omega < \omega_p \\ 0.09 & \omega > \omega_p \end{cases}, \quad (2)$$

which is converted to the wavenumber spectrum using the deep-water linear dispersion relation. In all simulations the peak wave period is $T_p = 2\pi/\omega_p = 10$ s; the corresponding peak wavenumber is $k_p = \omega_p^2/g \approx 0.040$ rad/m. The directional distribution is prescribed according to the $\cos^2$ function as

$$D(\theta) = \begin{cases} \dfrac{2}{\Delta\theta} \cos^2\left(\dfrac{\pi\theta}{\Delta\theta}\right), & |\theta| \leq \dfrac{\Delta\theta}{2} \\ 0, & |\theta| > \dfrac{\Delta\theta}{2} \end{cases}, \quad (3)$$

where $\theta$ is the angle with respect to the dominant wave direction (i.e., the $Ox$ axis). Every simulated sea state is characterized by the significant wave height $H_s$, spectral peakedness $\gamma$ and the directional spread $\Delta\theta$, see Table 1, and includes at least 7 realizations. Each realization of the initial condition is determined on the spatial domain of the size about 50 by 50 dominant wave lengths (about 8 km by 8 km) with the resolution $2^{10}$ by $2^{10}$ points. The surface velocity potential is determined according to the linear theory. The Fourier components of the initial condition are specified with randomized amplitudes and phases. The simulation starts with a nonlinear adaptation stage, when the nonlinear terms of the equations are slowly brought into play over $20T_p$, following the approach suggested by Dommermuth (2000). The consequent 1200 s ($120T_p$) of the wave evolution are used to produce the wave data for processing.

In the present simulations the wind effect is totally neglected, what may be regarded as the assumptions that it does not change the wave system significantly during the 20 minutes of the wave evolution. Only the wave conditions characterized by rare breaking events were simulated, what limits the maximum significant wave heights used for the modeling. The occasional wave breaking is regularized by the artificial hyperviscosity similar to the approach by Xiao et al. (2013). The imaginary part of the dispersion relation responsible for this operation is represented in the spatial Fourier domain for wave vectors **k** by the term

$$\Lambda(k) = \exp\left[-\left(\frac{|\mathbf{k}|}{\alpha k_p}\right)^{\beta}\right], \qquad (4)$$

with $\alpha = 16$, $\beta = 30$, what is a milder filter than in Xiao et al. (2013), where they used $\alpha = 8$. The resulting energy dissipation during the time span of the simulation is tiny: the total energy loss during a simulation is of the order of fractions of a percent or less. Thus, the simulations may be considered as conservative. The discussion of dealing with the occasional wave breaking may be found in [Slunyaev & Kokorina, 2020a].

Main parameters of the simulated sea states are collected in Table 1. The cases with significant wave height $H_s = 3.5$ m may be considered as ones with moderate wave steepness (experiment code A), whereas $H_s = 5…7$ m (codes B, C) correspond to strongly nonlinear sea states when the wave breaking starts to occur. As said above, larger $H_s$ lead to more frequent wave breakings, which cannot be described within the potential Euler equations. The peakedness factor $\gamma$ in the simulations is either 3 or 6 (upper index of the experiment code); the latter is frequently considered as more favorable for the Benjamin – Feir instability, as it corresponds to a narrower spectral peak. Sea states characterized by various directionality parameters (low index of the experiment code) have been modeled. The choices $\Delta\theta = 12°$ and $\Delta\theta = 62°$ are similar to the ones considered in Xiao et al. (2013) and correspond to relatively long crested and short crested waves respectively. The choice $\Delta\theta = 180°$ corresponds to the maximum spread allowed by the distribution (3), which should describe the so-called confused sea. Examples of the wave surfaces for four choices of the spreads $\Delta\theta$ at the terminal moments of simulations are shown in Fig. 2. The wave spectra do change during the wave evolution for 20 min, but typically not much, see in [Slunyaev & Kokorina, 2020b]. Therefore, in what follows we use the parameters of the initial wave spectra to characterize the simulated sea states.

Besides the spectral parameters, Table 1 contains the information about the nonlinear parameter $M$. Most of the simulations are performed for $M = 3$ to describe the 4-wave nonlinear interactions (experiment code B), though in addition some of the states were simulated with higher order of nonlinearity of the scheme, taking into account 5-wave interactions ($M = 4$, experiment code C), when the three-dimensional instability may become important and lead to specific pattern formation [McLean, 1982; Shrira et al., 1996]. Linear simulations have been also performed (experiment code L) for the purpose of validation of the statistical analysis procedures and evaluation of the broad-bandness effect.

## 3. Wave field decomposition approach

The role of specific nonlinear wave interactions and wave directions to the wave height PDF is analyzed in Sec. 4, where wanted wave components are extracted from the displacement field $\eta(\mathbf{x},t)$ using the spectral approach described in this section.

### 3.1. Windowed spatio-temporal Fourier transform

Consider the time frame of the duration $T_w$ with the mid instant $t$, see illustration in Fig. 3a. Then, the spatio-temporal Fourier transform is calculated as follows,

$$\hat{\eta}(\mathbf{k},\omega;t) = F_x F_y F_\tau (M(\tau-t)\eta(\mathbf{x},\tau)), \tag{5}$$

where $F_\xi(\cdot)$ denotes the Fourier transform with respect to the variable $\xi$, and $M(t) = 0.5\,(1 + \cos(2\pi t/T_w))$ is the smoothing Hann mask defined in the interval $[t - T_w/2,\ t + T_w/2]$. The arguments of the triple Fourier transform are the wave vector $\mathbf{k} = (k_x, k_y)$ and the cyclic frequency $\omega$. An example of isosurfaces of the Fourier amplitudes $|\hat{\eta}|$ for a particular time moment is given in Fig. 3b, where the wave vectors are presented in polar coordinates $(k, \theta)$, $k = |\mathbf{k}|$ and $\theta$ is the angle with respect to the $k_x$ axis. Only the part for $\omega > 0$ is displayed as the function $\eta$ is real-valued and hence $\hat{\eta}(-\mathbf{k},-\omega) = \hat{\eta}^*(\mathbf{k},\omega)$. Three leading orders of the Fourier amplitudes are shown, so that if the characteristic parameter of nonlinearity is 0.1, terms corresponding to three orders of nonlinearity should be visible in the plot.

Though the JONSWAP frequency spectrum is relatively broad and nonlinear harmonics essentially overlap in either frequency or wavenumber domains, the spectral lobes which may be associated with different nonlinear harmonics are well distinguishable in the wave-vector–frequency Fourier space, as follows from the examples in Fig. 3b and Fig. 4. Under the condition of a narrow spectrum including the particular limit of a single Stokes wave $(\mathbf{k}, \omega)$, the relation between frequencies and wave vectors of superharmonics are specified by the following formula:

$$\omega_n = n\Omega\left(\frac{\mathbf{k}}{n}\right) = \sqrt{ngk}\,, \qquad \Omega(\mathbf{k}) = \sqrt{gk}\,, \qquad k = |\mathbf{k}|, \tag{6}$$

which follows from the dispersion law $\omega = \Omega(\mathbf{k})$. Here $g$ is the acceleration due to gravity, and $n \geq 1$ is an integer which counts the nonlinear harmonics. In Fig. 4 the plots similar to Fig. 3b are given for four simulated rough sea states with different directional spreads, where, in contrast to Fig. 3b, the vertical axis represents the squared frequency. Respectively, the superharmonics for several $n$ are shown in the figure with dashed straight lines.

It is clear that in all examples given in Fig. 4, most of the spectral energy is concentrated approximately along the lines $\omega_n^2(k)$ with $n = 1$ (the fundamental wave component) and $n = 2, 3$, which may be associated with the second and the third nonlinear harmonics respectively. Besides, the induced low-frequency component is observed, which is located below the dispersion relation and is found to follow rather well the condition (6) with the choice $n = 0.5$ (see the lowest dashed lines in Fig. 4).

Based on the examples presented in Fig. 4, in order to select the $l$-th nonlinear harmonic, we introduce the spectral masks $W_l(\mathbf{k},\omega)$, $l = 0, 1, 2, 3$, which zeroize the Fourier amplitudes outside the following spectral areas: $0 \leq \omega < \omega_{0.75}$ for $l = 0$ (zeroth or difference harmonic); $\omega_{0.75} \leq \omega < \omega_{1.5}$ for $l = 1$ (first harmonic); $\omega_{1.5} \leq \omega < \omega_{2.5}$; for $l = 2$ (second harmonic) and $\omega_{2.5} \leq \omega < \omega_{3.5}$ for $l = 3$ (third harmonic). These filter boundaries are shown in Fig. 4 with dash-dotted lines; the selected spectral areas as the result of application of the masks $W_l$ are differentiated in Fig. 3b and Fig. 4 with the help of different colors, see the legends. One may conclude from the figures that the spectral filters used separate the isolated spectral lobes quite well in all the cases.

It is important that the spectral function $\hat{\eta}(\mathbf{k}, \omega; t)$ in Eq. (5) retains the complex phases. Then the wave harmonics $\eta_l(\mathbf{x}, t)$ may be obtained after application of the corresponding spectral filter $W_l(\mathbf{k}, \omega)$ and consequent triple inverse Fourier transform:

$$\eta_l(\mathbf{x};t) = F_x^{-1} F_y^{-1} F_t^{-1} \left( W_l \hat{\eta}(\mathbf{k}, \omega; t) \right). \tag{7}$$

Note that in order to obtain the two-dimensional surface displacement which corresponds to the *l*-th spectral area at a single instant of time $t_0$, the information about the original field $\eta(\mathbf{x}, t)$ is required over an extensive time interval $[t_0 - T_w/2, t_0 + T_w/2]$. In the present study we use the time frame of the duration $T_w = 256$ s, which corresponds to about 25 wave periods, with the resolution 0.5 s. According to the tests, this set of parameters provides both, a relatively good description of time sequences in physical space, and a sufficiently large Nyquist frequency, so that the aliasing effect in the frequency domain is not significant. The described circumstance makes the spectral filtering procedure costly in terms of time and use of computer memory.

In addition to separating different spectral areas representing particular nonlinear wave interactions, in this work we extract waves propagating in given angle cones. Waves moving in strictly opposite directions cannot be differentiated based on the spatial Fourier transform taken at a single time instant $t$, as $\hat{\eta}(-\mathbf{k}, t) = \hat{\eta}^*(\mathbf{k}, t)$. Owning the spatio-temporal transform, we have $\hat{\eta}(-\mathbf{k}, \omega; t) = \hat{\eta}^*(\mathbf{k}, -\omega; t)$, and thus the wave propagation angles within the full 360° can be resolved using the information about the wave frequency. Technically, it is convenient to calculate the direction angles of Fourier harmonics based on the product $\omega \mathbf{k}$; they coincide with the directions of corresponding phase velocities. The performance of this filter is illustrated in Fig. 5, where examples of the reconstructed surfaces are presented for the simulation $B_{62}{}^3$ with the directional spread $\Delta\theta = 62°$. Different panels exhibit waves traveling within certain intervals of direction angles (shown at tops of the panels as purple circle segments) with respect to the dominant wave direction (shown as black arrows). Waves in Fig. 5a look relatively long-crested, whereas the patterns in Fig. 5b resemble cross sea waves. Waves in Fig. 5c,d have less pronounced structures. In Fig. 5 the spectral masks $W_l$ for separating nonlinear wave harmonics were not applied. The color bar limits correspond to the maximum displacements of the constructed surfaces. One may see that waves traveling under larger angles to the dominant wave direction are characterized by smaller amplitudes, as expected.

### 3.2. Interpretation of the spectral data of nonlinear waves

The spectral plots like Fig. 3b provide rich information about the nonlinear wave fields. In particular, a small amount of counter propagating waves may be seen in Fig. 3b for the wave directions $\theta \approx \pm 180°$, though they were absent in the initial condition with the directional spread 62°. The opposite waves are produced in the course of the nonlinear wave evolution. They formally belong to the lobe of the first harmonic ($l = 1$) and are interpreted as such in what follows.

Considering an arbitrary spectrum, possibly broad in frequencies and direction angles, for any pair of waves ($\mathbf{k}_1, \omega_1$) and ($\mathbf{k}_2, \omega_2$) with $\omega_j = \Omega(\mathbf{k}_j)$, $j = 1,2$, the sum nonlinear harmonic ($\mathbf{k}_p, \omega_p$) = ($\mathbf{k}_1 + \mathbf{k}_2, \omega_1 + \omega_2$) and the difference nonlinear harmonic ($\mathbf{k}_m, \omega_m$) = ($\mathbf{k}_1 - \mathbf{k}_2, \omega_1 - \omega_2$) are generated in the leading order of nonlinearity (here the subscripts '$p$' and '$m$' code 'plus' and 'minus' respectively). According to the second-order nonlinear theory, the surface displacement $\eta(\mathbf{x},t)$ for the two waves with amplitudes $a_1$ and $a_2$ may be presented in the form

$$\eta(\mathbf{x},t) = a_1 \cos\psi_1 + a_2 \cos\psi_2 + \frac{a_1^2 k_1}{2}\cos(2\psi_1) + \frac{a_2^2 k_2}{2}\cos(2\psi_2) + \qquad (8)$$

$$+ a_1 a_2 B_p^\infty \cos(\psi_1 + \psi_2) + a_1 a_2 B_m^\infty \cos(\psi_1 - \psi_2),$$

$$\psi_j = \mathbf{k}_j \mathbf{x} - \omega_j t + \phi_j, \quad j = 1, 2,$$

where $\psi_j$ are the wave phases and $\phi_j$ are the reference phase constants. The third and fourth terms in (8) are the superharmonics of the principle waves. The combination terms in the second line of Eq. (8) are characterized by the pairs of sum and difference frequencies and wave vectors, ($\mathbf{k}_p, \omega_p$) and ($\mathbf{k}_m, \omega_m$). The deep water condition with the dispersion relation (6) is known to be unsuitable for triad resonances, thus $\omega_p \neq \Omega(\mathbf{k}_p)$ and $\omega_m \neq \Omega(\mathbf{k}_m)$. Considering the deep water limit, the interaction coefficients $B_p^\infty$ and $B_m^\infty$ in (8) depend only on the interacting wave vectors; the corresponding expressions may be found in [Dalzell, 1999]. These dependences are plotted in Fig. 6 as functions of the angle $\varphi$ between the wave vectors $\mathbf{k}_1$ and $\mathbf{k}_2$ for three wave length ratios.

As the magnitudes of the nonlinear interaction coefficients strongly depend on the angle $\varphi$, the spread of wave directions noticeably influences the picture of the spectral energy distribution. For waves with similar lengths (narrow frequency spectrum) the coefficients reach maximum values in the limits of either unidirectional waves, $B_p^\infty(\varphi = 0)$, or opposite waves, $B_m^\infty(\varphi = \pm 180°)$, see the red and blue broken lines in Fig. 6 correspondingly. The first situation corresponds to the generation of the second harmonic of a Stokes wave; the respective spectral lobes are well seen in Fig. 3b and Fig. 4. The second option should be obviously most relevant in simulations with the maximum directional spread, $\Delta\theta = 180°$, and corresponds to the generation of difference harmonics with small frequencies but not necessarily vanishing wavenumbers. In fact, these harmonics are not seen in Fig. 4d due to unsuitable scaling of the frequency axis. The same data as in Fig. 4d, but with the vertical axis indicating the wave frequency in the first power, are presented in Fig. 7 in two views. These graphs clearly show the energy spots of low-frequency waves (dark red) propagating in almost opposite directions $\pm 90°$ with wave vectors exceeding in length the carrier wavenumber $k_p \approx 0.04$ rad/m. According to the partitioning of the Fourier domain described in Sec. 3.1, they are treated as part of the zeroth (difference) harmonics which correspond to the spectral area with $l = 0$.

Note that the association of the spectral areas labeled with the indices $l$ (which specify the masks $W_l(\mathbf{k},\omega)$ as described in Sec. 3.1) with $l$-th nonlinear harmonics becomes relative in the case of a broad spectrum. The simple analysis performed in the Appendix leads to the conclusion that in general $\omega_p > \Omega(\mathbf{k}_p)$ and $\omega_m < \Omega(\mathbf{k}_m)$. Thus, in the wavenumber-frequency diagrams in Fig. 4 and Fig. 7 the quadratic sum and difference harmonics are always located above and below the dispersion relation, respectively. It may be further shown (see the Appendix), that the sum frequencies in the $(\mathbf{k},\omega)$ plain can lie well above the second nonlinear harmonic $\omega_2(\mathbf{k}) = (2gk)^{1/2}$ but below the function $\omega_{n_p}(\mathbf{k})$, where the parameter $n_p$ is specified by the directional spread, see Eq. (A.3) and Fig. 8. The sum frequencies can also occupy the area below the second harmonic $\omega_2(\mathbf{k})$ but above the linear dispersion relation $\Omega(\mathbf{k})$, where the low boundary is determined by the diversity in wave lengths (frequency spectrum width), see Fig. 8. Some conclusions about the location of the difference harmonic can be also drawn (see the Appendix). In particular, waves with broader angular spectrum can generate lower frequencies, though the upper limit of the difference harmonic is controlled by the low-frequency part of the spectrum.

The representation of higher order nonlinear combination harmonics in the form similar to (8) is also possible [Madsen & Fuhrman, 2006], but is more complicated. Correspondingly, the analysis of the Fourier domain becomes much more difficult due to the mixing of different terms in the same spectral areas. Besides, quartet (and higher order) nonlinear wave resonances become allowed; significant deviations from the linear dispersion relation may occur due to the nonlinear frequency shift and formation of persisting coherent wave structures. Thus the interpretation of spectral regions in the situation of strongly nonlinear simulations is significantly less certain. Nevertheless, the introduced spectral filters which operate in the spatio-temporal Fourier space are considered to be a powerful tool for comprehensive analysis of directional wave systems. Based on the linear property of these transforms, different combinations of the selected harmonics can be considered. This approach was used in [Slunyaev, 2020] to evaluate the dynamic kurtosis and in [Slunyaev, 2023] to estimate the contribution of different nonlinear wave harmonics to the wave height probability distributions. It is used in the next section to analyze the statistics for down- and up-crossing waves.

## 4. Rogue wave probability

The vast majority of sea wave measurements are represented by single point records. The contemporary approaches in oceanography are based on the analysis of discrete time series obtained by various instruments. In particular, the generally accepted concept of an individual wave is formulated for a time series, as described in the Introduction, and most of the mathematical toolkit is based on this representation. Though having the information about 2D water surface evolution, in this work we aim to follow conventional approaches and thus investigate the wave height properties based on the zero-crossing analysis of the surface displacement data series.

In Sec. 4.1 below we consider the surface elevation time series retrieved at fixed locations in the simulated waves domain. The spatio-temporal spectral filtering described in Sec. 3 requires much computational efforts but should be performed for each instant of time. In order to reduce the computational cost, the analysis of contribution of different nonlinear wave harmonics presented in Sec. 4.2, is conducted for space series obtained for a sequence of time instants. Otherwise, the analysis of time sequences and space sequences is performed in the same way. In Sec. 4.3 the analysis of the space series is further detailed by considering different directions of wave propagation.

**4.1. Time series analysis**

The databank of time series of the surface displacement consists of sequences $\eta_{ij}(t)$ collected at every node $x_i$ of the coordinate grid along the $Ox$ axis ($1 \leq i \leq 2^{10}$) and 64 equidistant slices $y_j$ of the transverse coordinate. Thus, the configuration of the 'measurement' locations looks like the set of equidistant lines parallel to the dominant wave propagation direction, see the red secant planes in Fig. 3a. The actual spacing between the secant planes is four times smaller than shown in the figure. By collecting the data from a large area of space, we aim to obtain a richer statistical ensemble. At the same time, the sequences taken at close points in space may be not completely independent statistically. This issue was discussed in [Slunyaev & Sergeeva, 2013] in application to the simulation of unidirectional waves; it was shown that: (i) the exceedance probability distribution for linear and weakly nonlinear waves follow the Rayleigh distribution (1) rather well, when the significant wave height is calculated as one-third of the highest waves in the series, $H_s = H_{1/3}$; and (ii) the correlation distance between time sequences is significantly reduced when the nonlinearity grows. Therefore, one may hope that the effect of imperfect statistical independence of the processed data does not lead to noticeable artifacts.

Two examples of the comparison between the exceedance probability distributions calculated for waves simulated within the linear theory for the cases of long-crested and short-crested waves are given in Fig. 9a and Fig. 9b respectively. Note that the horizontal axis in the plots represents the squared normalized heights (or amplitudes $A$), and the thick green solid line is the theoretical Rayleigh distribution (1). The use of $H_s = H_{1/3}$ instead of $H_s = 4\sigma$, where $\sigma$ is the root-mean-square surface displacement, leads to the extension of the interval of agreement with the Rayleigh law to larger wave events (compare the thick and thin lines for $H_{up}$ and $H_{down}$). The distributions for the crest and trough amplitudes fit the Rayleigh law even better (for normalization, we define $A_s = 2\sigma$), probably because they do not imply the narrow-band approximation, which is poorly satisfied by the JONSWAP spectrum. One can also see that the simulated waves agree with the theoretical distribution (1) noticeably better when the directional spread grows (cf. Fig. 9a and Fig. 9b). This may be related to the fact that under the condition of a broader spectrum (either in wavenumbers or angles) a greater number of Fourier modes hold significant amount of energy, what is a favorable situation for the occurrence of constructive wave interference, which is the only mechanism of large wave generation in the linear framework. The PDFs for simulated waves drop down in the range of the most extreme waves due to the finite number of wave samples. It may be concluded that the Rayleigh distribution is indeed a reasonable first approximation for the simulated waves. No noticeable difference between up-crossing and down-crossing waves, and also between wave crest and wave trough amplitudes ($A_{cr}$ and $A_{tr}$ respectively) is found within the linear approximation.

The wave nonlinearity leads to modification of the PDFs, as discussed in a large amount of literature, and may result in a noticeable increase in the probability of large wave generation if the spectral width is relatively small (e.g., [Onorato et al, 2001, 2009; Janssen, 2003] among many others). The discussion of this effect based on the wave databank used in the present study may be found in [Slunyaev & Kokorina, 2023]. Below we consider the effect of nonlinearity on the emerging difference between the probabilities of generation of very large down-crossing and up-crossing waves.

As shown below, the simulations of nonlinear waves exhibit wave height probability distribution functions, which may be noticeably different in the interval of extreme events, depending on whether the up-crossing ($P_{up}$) or down-crossing ($P_{down}$) method is used to select individual waves and corresponding heights. To estimate this difference, we compare the exceedance probabilities calculated from the simulated data for the given amplification factors $AI = H/H_s$, where $H_s = 4\sigma$ is used hereafter for the sake of simplicity. In Fig. 10, the fractions of down-crossing waves (i.e., of the type shown in Fig. 1a) in the sets of waves with $AI$ exceeding the values 1.8 (green bar), 2.0 (pink bar) and 2.2 (yellow bar) are shown for the simulations from Table 1. The portions are calculated as $P_{down}/(P_{down} + P_{up})$; the horizontal dotted line at 0.5 represents the reference of the parity between up- and down-crossing wave probabilities. The horizontal axis in Fig. 10 indicates different angular widths $\Delta\theta$ from 5° to 180°. The size of the symbols denotes the strength of nonlinearity (i.e., the parameter $H_s$); the black small triangles correspond to linear simulations. The downward and upward triangles correspond respectively to the cases of small and large peakedness factors. The blue colored symbols code the simulations with the nonlinearity parameter $M = 3$ whereas the red color corresponds to the simulations with higher order of nonlinearity $M = 4$.

One can see from Fig. 10 that in the sea states characterized by large $H_s$ (big triangles) the amount of huge down-crossing waves becomes smaller compared to up-crossing waves, hence rear rogue waves like shown in Fig. 1b dominate. The difference between probabilities of down- and up-crossing waves systematically grows with the exceedance $AI$. It is most pronounced for the intermediate values of the width of the angular spectrum of 62° and 90°, where the fraction of down-crossing waves (i.e., frontal waves shown in Fig. 1a) with $H/H_s >$ 2.2 drops down to about 25%; the difference is almost absent in the case $\Delta\theta = 5°$. Note that in the latter case the significant wave height is limited by the value of 5 m, what is smaller than in other simulations due to seemingly more significant wave breaking effect. The difference between the number of frontal and rear rogue waves was found noticeable in the simulations of unidirectional waves in [Slunyaev & Sergeeva, 2013] and [Slunyaev et al., 2016] with $H_s =$ 7 m, where another strategy for regularization of the wave breaking was used. According to Fig. 10, the balance between frontal and rear rogue waves approximately holds for any directional spread under the condition of relatively small waves with $H_s = 3.5$ m.

The simulations with different peakedness parameters (see orientation of the triangles) were conducted for the directional spreads 12° and 62°; they show little difference between $P_{down}$ and $P_{up}$ for the largest amplification factor $AI > 2.2$. The difference seems to be slightly stronger for small $\gamma$, what is probably a not trustworthy result due to the relatively small number of wave samples. The number of individual waves in the subsets for different simulations presented in Fig. 10 are at least 34 000 for $AI > 1.8$, at least 4 500 for $AI > 2$ and between 230 and 30 000 for $AI > 2.2$.

The simulations with different nonlinear parameters $M = 3$ and $M = 4$ (coded by the color of the triangles) demonstrate almost identical results in terms of the balance between the up-crossing and down-crossing large wave probability.

## 4.2. Space series analysis: Contributions of nonlinear harmonics

In order to identify the role of particular nonlinear wave interactions in the discovered disproportion in the number of frontal and rear rogue waves, we consider the PDFs for the surface wave fields which include only certain wave harmonics, when the unwanted ones are filtered out using the method described in Sec. 3.1. Since the spatio-temporal filtering is a time consuming procedure, instead of dealing with time series, in this section we analyze space series $\eta_{jq}(x)$, where the index $j$ corresponds to 64 transverse locations as described in Sec. 4.1 and the index $q$ counts 2 400 time points of the "raw data" in the interval 200 s $\leq t \leq$ 1400 s. The windowed Fourier procedure is applied to 59 overlapping time intervals from the analyzed period with the duration $T_w = 256$ s and the stepping 16 s. Thus, the transformed data contains 64×59 8-km space samples per each realization.

At first, we repeat the analysis presented in Fig. 10, but now process the raw space series, see Fig. 11. In the space series the rear rogue waves are down-crossing waves, see Fig. 1b. Respectively, in Fig. 11 we give the fractions of up-crossing waves in space series, to be consistent with Fig. 10 which characterizes time series. In other respects, the ways of presenting the results in Fig. 10 and Fig. 11 are identical. The number of individual waves in the space series subsets are at least 17 000 for $AI > 1.8$, more than 1 400 for $AI > 2$ and between 13 and 14 000 for $AI > 2.2$. The minimum number of waves corresponds to the linear simulation with the directional spread $\Delta\theta = 12°$; in Fig. 11 it is represented by an outlier in the range $AI > 2.2$ (see the small black symbol on the yellow bar above the level of 0.6).

One can see that Fig. 10 and Fig. 11 are fully consistent. At the same time, the difference between portions of frontal and rear extreme waves is noticeably larger in the space series. In the wave snapshots the fraction of frontal rogue waves with $AI > 2.2$ under the conditions $H_s = 7$ m and $\Delta\theta = 62°$ falls below 20%.

In Fig. 12 we present the ratios $P_{up}/(P_{down} + P_{up})$ as functions of the wave height excess $H/H_s$, $H_s = 4\sigma$. These ratios are calculated for the original surface displacements ('Full') and for different combinations of wave components labeled according to the harmonic numbers $l$ as described in Sec. 3 (see the legend). The cases displayed in Fig. 12a-e correspond to the selection of the most intense sea states simulated for 5 spread angles from 5° to 180° with small peakedness (except the case $\Delta\theta = 5°$ in Fig. 12a), see the shaded rows in Table 1. The simulations with the nonlinear parameter $M = 4$ are shown (except for Fig. 12a) as most reliable. Any data point shown in Fig. 12 corresponds to at least 50 individual waves, whereas the data points based on over 500 individual waves are given by thicker lines. One can see that the maximum wave amplification $H/H_s$ is very different in the considered examples. It reaches the highest value in the case of a narrow angle spectrum (Fig. 12a) and is significantly smaller in the simulations of short-crested and confused seas (Fig. 12c-e), what is consistent with the general view on the rogue wave probability in directional seas [Onorato et al., 2009]. One can also clearly see that the portion of frontal large waves decreases gradually when $H/H_s$ grows (see the green line for the full displacement), what is a particularly strong effect for the directional spread $\Delta\theta = 62°$ (Fig. 12c).

Similar estimations of the fractions of frontal extreme waves in the fields which correspond to solely the first harmonic are given in Fig. 12 with cyan lines (see '1' in the legend). If the wave phases are uncorrelated, it must show the balance between up- and down-crossing waves and follow the horizontal dotted line. Some deviations from this reference are observed for the most extreme waves (particularly in Fig. 12c), though the first harmonic exhibits the disproportion inverse to the full field case, so that the probability of frontal rogue waves is up to 5% larger when calculated for at least 500 waves.

The combination of the first and zeroth harmonics (see the blue curves for '0+1') yields the inverse disproportion between the up- and down-crossing waves as well. The description is improved when the second harmonic is taken into account in addition (see the red curves for '0+1+2'). A relatively accurate but somewhat exaggerated description of the balance between up- and down-crossing waves is provided by the combinations of just the first and the second harmonics ('1+2'). A very accurate result is reproduced only in the case of simultaneous consideration of harmonics from the zeroth to third (black curves, '0+1+2+3'). We should note that the results are rather uncertain in the cases $\Delta\theta = 5º$ (Fig. 12a) and $\Delta\theta = 180º$ (Fig. 12e), but look consistent with the other cases.

### 4.3. Space series analysis: Investigation of dependence on the wave direction

Thus, the difference between probability of frontal and rear extreme waves does not come from the spectral areas below the dispersion law, but is determined by the spectral area $l = 2$ associated with the second harmonic. Since the nonlinear interaction coefficients may strongly depend on the angle between interacting waves (see Fig. 6 for the quadratic nonlinearity), and the asymmetry effect is found to require sufficiently broad angular spectrum, it is worthy to examine the dependence of the concerned wave asymmetry on the wave direction angle with respect to the main wave course. Do oblique waves contribute the most to the effect? In this subsection we analyze the difference between down- and up-crossing heights of waves propagating within certain direction angle intervals, using the spectral filter discussed in Sec. 3.1 and demonstrated in Fig. 5. The strongly nonlinear simulation $C_{62}^6$ is used as a representative example, which exhibits the most significant difference between the statistics for down-crossing and up-crossing waves (see Fig. 12c).

The functions $P_{up}/(P_{down} + P_{up})$ are calculated from the space series of waves propagating under the angles with respect to the main wave direction which belong to the given intervals $\theta \in [-\Theta, \Theta]$. The functions are plotted in Fig. 13 for five values of $\Theta$ starting from $10°$ (see the legend), where the case $\Theta = 180°$ corresponds to the full range of angles; the spectral filters $W_l$ were not applied. For smaller angles $\Theta$ the maximum wave height in the wave data reduces, and the difference between $P_{up}$ and $P_{down}$ vanishes consequently. It follows from Fig. 13 that the difference between probability of very high frontal and rear waves starts to appear from relatively small direction spreads 15 and quickly converges to the curve for the full field. The dependence does not demonstrate noticeable variation with $\Theta$.

Similar functions $P_{up}/(P_{down} + P_{up})$ for waves propagating within sectors of different directions are plotted in Fig. 14, but now the spectral filter $W_1$ is applied which selects the spectral area in the spatio-temporal Fourier domain associated with the first harmonic. Though the tails of the curves in Fig. 14 which correspond to the maximum waves do not follow the reference balance value 0.5 perfectly, the deviations are too small to conclude confidently about a difference between $P_{up}$ and $P_{down}$ for any angular spread $\Theta$. However, all the curves seem to agree with the result for the full range of angles $\Theta = 180°$ and do not exhibit a clear dependence on $\Theta$.

Note that a slight but distinct deviation from the balance between up- and down-crossing waves is observed in Fig. 13 in the interval of relatively small heights $H/H_s < 0.5$, see the inset. Nothing of the kind is observed in the case of the first harmonic shown in Fig. 14.

## 5. Discussion

In this work, with the help of the direct numerical simulation of potential equations of hydrodynamics, we investigate a new asymmetry of nonlinear waves, which manifests itself through the disproportion between the number of frontal and rear rogue waves. This asymmetry leads to different estimations of the rogue wave probability according to the down- and up-zero-crossing analyses of wave records. The difference in depths of preceding and following troughs of the average rogue wave shape constructed from the observational data may be found in Cattrell et al. (2018) (though they seem to report the opposite asymmetry than we do). Pinho et al. (2004) reported remarkably greater number of up-crossing than down-crossing rogue waves in their instrumental time series. Similar to Pinho et al. (2004), more probable rear rogue waves were found in the direct numerical simulations by Xiao et al. (2013) and in our previous works [Slunyaev & Sergeeva, 2013; Slunyaev et al., 2016; Slunyaev & Kokorina, 2020]. At the same time, such wave asymmetry was not observed within the equations for wave modulations which employed the narrowband assumption (the nonlinear Schrödinger equation and its modified version by Trulsen & Dysthe (1996)).

The present study of directional irregular deep-water waves with single-peaked JONSWAP spectra shows that under appropriate conditions, rear rogue waves may be remarkably more likely than rogue waves with deeper troughs in front, corresponding to a higher probability of up-crossing rogue waves in time series. The conditions favorable for the observation of this effect are strong nonlinearity and relatively broad angular spectrum (rough short-crested seas). The effect is significant only when the average wave steepness becomes sufficiently large, $k_p \sigma \approx 0.05$, while it is negligible when $k_p \sigma \approx 0.035$ and below. The difference between probabilities of down-crossing and up-crossing rogue waves reduces in the limits of paraxial waves and confused seas with very broad directionality. Based on the performed analysis, the peakedness factor $\gamma$ taken in the interval from 3 to 6 plays very little role if any. The difference in probabilities strengthens for larger wave amplifications $H/H_s$. It is most significant for zero-crossing analysis of space series (i.e., surface snapshots) rather than traditional time series, and may reach the proportion 5 to 1 for $H/H_s > 2.2$. Thus, the rogue wave probability estimate can be strongly influenced by the choice of either down- or up-crossing method when analyzing the wave data.

Very steep wave events are accompanied by the breaking, which is regularized in this work by introducing a hyperviscosity term (4) into the numerical scheme. The controlling parameters of the viscosity were carefully selected in [Slunyaev & Kokorina, 2020] with the goal to introduce as small perturbation to the wave evolution as possible. However, this term has a greater effect on the most extreme waves, and hence potentially may distort the rogue wave picture. A dedicated numerical simulation was conducted in order to clarify the effect of the artificial hyperviscosity on the difference between the probability distributions for down- and up-crossing waves. This simulation was performed for the same conditions as in the experiment $B_{62}3$, but with noticeably stronger viscosity taking $\alpha = 10$ in (4). It was verified that the difference between PDFs for down- and up-crossing waves in the simulation with stronger viscosity was very close but slightly smaller than in the original simulation when plotted in Figs. 10 and 11. Therefore, the regularization of occasional breaking is unlikely to be the reason of the difference between the probabilities of large down- and up-crossing waves in our simulations.

The revealed difference between troughs preceding and following rogue waves is identified in both, space- and time-series. In terms of the Fourier transform, this geometric feature is determined by harmonics with scales exceeding the length and period of the dominant wave. Generally speaking, it could be a result of two kinds of nonlinear effects: (i) the coherent wave dynamics which can support persistence of skewed wave modulations (with zero mean) or/and (ii) the phase-locked bound waves, where the most obvious candidate is the induced long-scale and low-frequency displacements which can deform the waves travelling overhead.

Thanks to the Central Limit Theorem, independent random waves obey the Gaussian statistics and do not exhibit on average any asymmetry of extreme events with respect to the vertical line. The second order nonlinear solution is given by (8). Since only non-resonant 3-wave interactions can occur in deep water, the coherent wave dynamics is not expected to occur due to the quadratic nonlinear effects. The set-down of wave groups with narrow spectrum follows from the second-order theory for the difference harmonic [Adcock & Taylor, 2009] and is effectively described by the Dysthe model for wave modulations [Dysthe, 1979; Trulsen & Dysthe, 1996], in which it is quadratic in nonlinearity, but corresponds to the third order of the asymptotic theory due to the additional assumption of weak dispersion.

Within the Dysthe theory, the leading-order nonlinear wave harmonics which constitute the surface displacement $\eta(\mathbf{x},t)$ are given by [Trulsen, 2007; Slunyaev, 2005]:

$$\eta = \eta^{(0)} + \eta^{(I)} + \eta^{(II)} + \eta^{(III)} + ..., \qquad (9)$$

$$\eta^{(I)} = \mathrm{Re}\left(A e^{i\omega_0 t - i k_0 x}\right), \quad \eta^{(II)} = \frac{k_0}{2} \mathrm{Re}\left(A^2 e^{2i\omega_0 t - 2i k_0 x}\right),$$

$$\eta^{(III)} = -\frac{1}{4} \mathrm{Im}\left(\frac{\partial}{\partial x}(A^2)\, e^{2i\omega_0 t - 2i k_0 x}\right) + \frac{3k_0^2}{8} \mathrm{Re}\left(A^3 e^{3i\omega_0 t - 3i k_0 x}\right), \quad \eta^{(0)} = \frac{1}{4} H\left\{\frac{\partial}{\partial x}|A|^2\right\},$$

where $A(\mathbf{x},t)$ is the complex envelope for the surface displacement and $H(\cdot)$ denotes the Hilbert transform, which is an odd function of its argument.

From the form of the solution for the mean displacement $\eta^{(0)}$ (9), it can be understood that it inherits the shape of the modulation of the dominant wave harmonic (free wave) $\eta^{(I)}$. If the face and rear sides of free wave groups are symmetric (what is true on average within, for instance, the linear theory or the cubic nonlinear Schrödinger equation), then the induced by the quadratic nonlinearity set-down preserves the group symmetry.

The expression for the quadratic in nonlinearity second harmonic $\eta^{(II)}$ cannot contribute to the asymmetry of wave groups with respect to the vertical line. For particular envelopes, the third harmonic $\eta^{(III)}$ may cause asymmetry of the total displacement $\eta$ with respect to the vertical line due to the mixture of two summands. But this asymmetry should vanish after an averaging, assuming that the complex phases of $A$ are random. Thus, there is no source of the wave group asymmetry with respect to vertical line within the second-order theory for slowly modulated fully uncorrelated waves.

Real-world wave groups are known to be asymmetric with steeper frontal and smoother rear sides, see e.g. Shemer et al. (2001), what seems to be qualitatively consistent with the domination of rear rogue waves. This 'triangular' asymmetry is reproduced by the Dysthe model (e.g. [Trulsen, 2007]) due to the terms of nonlinear dispersion entering the evolution equation, which are cubic in terms of the wave steepness. Based on the numerical simulation of disintegration of Gaussian wave envelopes within the Dysthe equation, it was shown in [Slunyaev, 2015] that the wave group asymmetry may be in fact different depending on the amplitude of the initial perturbation. At the same time, the difference between statistics for down-crossing and up-crossing waves within the stochastic simulation of the Dysthe equation has not been reported so far. Though the long-scale term $\eta^{(0)}$ in the classic Dysthe theory (9) does not possess its own asymmetry, higher-order expansions for the mean flow have more complicated structures [Sedletsky, 2003; Slunyaev, 2005] and theoretically can contribute to the wave group skew.

Using the spectral decomposition of the surface displacement into different nonlinear harmonics, we have examined the contribution of the Fourier region associated with the fundamental harmonic to the considered asymmetry. The basic assumption of the Gaussian statistics is violated if free waves become coherent. Some weak effect is indeed found, which however yields the difference in PDFs inverse to the net effect. Note that in [Slunyaev, 2020] the same filtering in the spatiotemporal Fourier domain was able to reveal the wave phase coherence leading to a significant increase of the dynamic kurtosis under certain conditions. The accounting for the low frequency part of the Fourier domain below the surface $\Omega(\mathbf{k})$ representing the dispersion relation, which is associated with the induced long-scale displacements, makes the imbalance between down- and up-crossing large waves more substantial, but again leads to an inverse disproportion between the PDFs.

The qualitatively correct balance between probabilities of down- and up-crossing waves is obtained only when high-frequency (relative to the dispersion law $\Omega(\mathbf{k})$) nonlinear harmonics are taken into account. An accurate description of the difference in PDFs is provided when the spectral areas associated with three orders of the nonlinear combination harmonics and the difference harmonic are considered together.

The deviation between probabilities of down- and up-crossing waves is beyond the second-order theory and appears in the simulations of broad-banded waves which account for up to 4-wave nonlinear interactions. Accounting for even higher order of nonlinearity, up to 5-wave interactions, does not change the situation noticeably. Though the effect is most substantial for waves with broad angular spectrum, it is not primarily driven by oblique waves. We conclude that the discovered domination of rear rogue waves (in other words, the higher probability of up-crossing waves in time series) is not a direct result of only coherent dynamics of free waves or the phase-locked induced mean flow, but rather is the result of their combination, which is mainly third-order in nonlinearity. The analytic theory able to describe this effect is absent so far.

**Appendix. Location of the combination harmonics due to the quadratic wave interactions in the spatio-temporal Fourier domain.**

Due to the nonlinearity of the hydrodynamic equations, each pair of waves (**k**$_1$,ω$_1$) and (**k**$_2$,ω$_2$) yields the sum ('plus') and difference ('minus') nonlinear wave harmonics (**k**$_p$, ω$_p$) = (**k**$_1$+**k**$_2$, ω$_1$+ω$_2$) and (**k**$_m$, ω$_m$) = (**k**$_1$−**k**$_2$, ω$_1$−ω$_2$) respectively. For the deep water gravity wave dispersion relation ω = Ω(**k**), where Ω(**k**) = Ω(k) = (gk)$^{1/2}$ and k = |**k**|, we introduce the parameters $n_p$ and $n_m$ as follows,

$$n_p = \left(\frac{\omega_p}{\Omega(k_p)}\right)^2, \quad n_m = \left(\frac{\omega_m}{\Omega(k_m)}\right)^2. \quad (A.1)$$

The numbers $n_p$ and $n_m$ correspond to the parameter $n$ in Eq. (6). For a given $n$ the corresponding dependence reads $\omega_n(\mathbf{k}) = n^{1/2}\Omega(\mathbf{k}) = (ngk)^{1/2}$, see (6). Hence, $n_p$ and $n_m$ specify the locations of combination harmonics in the domain (**k**, ω) with respect to the linear dispersion curve Ω(**k**), and also in the plane (k, ω) with respect to the curve Ω(k), thanks to the isotropy of the wave dispersion. The functions ω = $\omega_n$(**k**) for integer numbers $n$ = 2,3,… describe the Stokes wave superharmonics.

After straightforward calculations, one can obtain that

$$n_p^2 = \frac{f + 4h + 6}{f + 2\cos\varphi}, \quad n_m^2 = \frac{f - 4h + 6}{f - 2\cos\varphi}, \quad (A.2)$$

$$\text{where} \quad f(k_1, k_2) = \frac{k_1}{k_2} + \frac{k_2}{k_1}, \quad h(k_1, k_2) = \sqrt{\frac{k_1}{k_2}} + \sqrt{\frac{k_2}{k_1}},$$

and $\varphi$ is the angle between vectors **k**$_1$ and **k**$_2$. The constrains on the numbers $n_p$ and $n_m$ may be found based on the property $2 \le h \le f$ which holds for any positive $k_1$ and $k_2$, and assuming that the spread of angles $\varphi$ between the interacting waves is limited from above, $|\varphi| \le \Delta\theta \le 180°$. The constraints on $n_p$ and $n_m$ are further tightened when a finite band of wavenumbers is allowed due to the finite width of the frequency spectrum.

For the sum harmonics (**k**$_p$, ω$_p$), their frequencies $\omega_p = n_p^{1/2}\Omega(\mathbf{k}_p)$ are bounded within the limits specified by the allowed numbers $n_p$:

$$1 < n_p \le \frac{2}{\cos\dfrac{\Delta\theta}{2}}. \quad (A.3)$$

The upper limit in (A.3) tends to infinity when waves propagate in all directions ($\Delta\theta$ = 180°). It corresponds to the third harmonic of a Stokes wave $\omega_3$(**k**) for the angular spread $\Delta\theta \approx 96°$, and corresponds to the second harmonic of a Stokes wave $\omega_2$(**k**) in the case of unidirectional waves $\Delta\theta$ = 0. The choice $n$ = 2.5, which is used in Sec. 3.1 as the upper boundary for the spectral area $l$ = 2 which is associated with the second harmonic, corresponds to the directional spread $\Delta\theta \approx 74°$.

The lower limit in (A.3) is specified by interactions between waves with different lengths (which are allowed to be any in (A.3)). If the frequency spectrum is finite in width, the lower bound grows. In the particular limit of waves with equal frequencies (which however can travel under all angles), the limits (A.3) modify to the following:

$$2 < n_p \leq \frac{2}{\cos\frac{\Delta\theta}{2}}, \qquad (A.4)$$

hence the lower bound tends to the second harmonic of a Stokes wave. The upper limit in (A.4) reduces to the value of 2 under the restriction $\Delta\theta = 0$ (uniform Stokes wave).

So, according to the simple analysis above, the combination harmonics $(\mathbf{k}_p, \omega_p)$ in the situations of directional wave spectra with finite frequency and direction bands are expected to contribute to the areas of the Fourier domain in the vicinity of the second-order superharmonic $\omega_2(\mathbf{k}) = 2^{1/2}\Omega(\mathbf{k})$. Roughly speaking, the energy above it is brought by interactions between waves with different directions ($2 < \max(n_p) < \infty$), whereas the energy below $\omega_2(\mathbf{k})$ is due to the interactions between waves with different lengths ($1 < \min(n_p) < 2$), see the blue area in Fig. 8.

The analysis of the difference harmonic $(\mathbf{k}_m, \omega_m)$ can be performed in a similar way, what under the assumption $f \neq 2$ gives

$$0 \leq n_m < \frac{1}{1 + 4\frac{\sin^2(\Delta\theta/2)}{f-2}}. \qquad (A.5)$$

For narrow angle spectra ($\Delta\theta \to 0$) or very broad frequency spectra ($f \to \infty$) the upper limit in the relation (A.5) reduces to the dispersion relation:

$$0 \leq n_m < 1. \qquad (A.6)$$

In the situation of directional waves with a finite width of the frequency spectrum the upper bound of $n_m$ is smaller. It is clear that when the two wave components are characterized by similar frequencies (the case $f \approx 2$), the difference frequency is about zero, and $n_m \approx 0$. Then, the difference wavember may be not small if the angular spectrum is broad, what takes place in Fig. 7 with $\Delta\theta = 180°$.

The case of waves with simultaneously narrow frequency and angle spectra, $\mathbf{k}_1 \approx \mathbf{k}_2$, is specific. In this limit one can write

$$\frac{\omega_m}{\mathbf{k}_m} \approx \nabla_\mathbf{k}\Omega(\mathbf{k}_1) = \mathbf{C}_{gr}(\mathbf{k}_1), \qquad (A.7)$$

where the vector $\mathbf{C}_{gr}$ is the group velocity calculated for the wave $(\mathbf{k}_1, \omega_1) \approx (\mathbf{k}_2, \omega_2)$. Then, the location of the difference harmonic in the spectral domain for a given $\mathbf{k}_1$ will be represented by the line

$$\omega_m(\mathbf{k}) \approx \mathbf{k}\mathbf{C}_{gr}(\mathbf{k}_1) = \frac{\omega_1}{2k_1}k\cos\varphi, \qquad (A.8)$$

where $\varphi$ is the angle between vectors $\mathbf{k}_1$ and $\mathbf{k}$, as before Therefore, according to (A.8) the spread of inclinations of the linear functions $\omega_m(k)$ is determined by the spread of wave vectors $\mathbf{k}_1$ in both, absolute values and directions. The right-hand-side of Eq. (A.8) is confined from above by max($|\mathbf{C}_{gr}|$)·$k$. For the water wave dispersion law this upper limit of group velocities is specified by the low-frequency bound of the spectrum (see the dark-red color region below the dispersion curve in Fig. 8). The relation $\omega_m(k)$ is confined from below by min($|\mathbf{C}_{gr}|$)cos($\theta$)·$k$ if $\Delta\theta < 90°$, and may fall down to zero for greater spreads $\Delta\theta$. In the Fourier domain the energy of the difference harmonic of narrow banded waves with the dominant wave vector $\mathbf{k}_0$ should be located close to the origin along the straight line $\omega_m = |\mathbf{C}_{gr}(\mathbf{k}_0)|k = \omega(k_0)/k_0/2$. Beyond the narrow-banded approximation, the linear dependence on $k$ in (A.8) breaks, and the difference harmonic lobe becomes curved.

**Acknowledgements**

AS acknowledges the support of the Laboratory of nonlinear hydrophysics and natural hazards of V.I. Il'ichev Pacific Oceanological Institute, Far Eastern Branch Russian Academy of Sciences, project of Ministry of science and education of Russia, agreement No 075-15-2022-1127 from 1 July 2022, and the support from the Foundation for the Advancement of Theoretical Physics and Mathematics "BASIS", project No 22-1-2-42. The authors acknowledge the support by the state contract No 0030-2021-0007.

**Data Availability**

The data that support the findings of this study are available from the author upon reasonable request.

Table 1. Parameters of the numerical simulations. The coding: L – linear simulation; A – moderate steepness, $M = 3$; B – large steepness, $M = 3$; C – large steepness, $M = 4$. The subscripts denote $\Delta\theta$, the superscripts denote $\gamma$.

| Experiment code | Width of the angular spectrum, $\Delta\theta$, degrees | Peakedness $\gamma$ | Significant wave height $4\sigma$, m | Nonlinear parameter of the code, $M$ ($n$-wave interactions incl.) |
|---|---|---|---|---|
| $B_5^6$ | 5 | 6 | 5 | 3 (4-wave) |
| $L_{12}^3$ | 12 | 3 | 7 | linear |
| $A_{12}^3$ | 12 | 3 | 3.5 | 3 (4-wave) |
| $B_{12}^3$ | 12 | 3 | 6 | 3 (4-wave) |
| $C_{12}^3$ | 12 | 3 | 6 | 4 (5-wave) |
| $A_{12}^6$ | 12 | 6 | 3.5 | 3 (4-wave) |
| $B_{12}^6$ | 12 | 6 | 6 | 3 (4-wave) |
| $C_{12}^6$ | 12 | 6 | 6 | 4 (5-wave) |
| $L_{62}^3$ | 62 | 3 | 7 | linear |
| $A_{62}^3$ | 62 | 3 | 3.5 | 3 (4-wave) |
| $B_{62}^3$ | 62 | 3 | 7 | 3 (4-wave) |
| $C_{62}^3$ | 62 | 3 | 7 | 4 (5-wave) |
| $A_{62}^6$ | 62 | 6 | 3.5 | 3 (4-wave) |
| $B_{62}^6$ | 62 | 6 | 7 | 3 (4-wave) |
| $C_{62}^6$ | 62 | 6 | 7 | 4 (5-wave) |
| $L_{90}^3$ | 90 | 3 | 7 | linear |
| $A_{90}^3$ | 90 | 3 | 3.5 | 3 (4-wave) |
| $B_{90}^3$ | 90 | 3 | 7 | 3 (4-wave) |
| $C_{90}^3$ | 90 | 3 | 7 | 4 (5-wave) |
| $L_{180}^3$ | 180 | 3 | 7 | linear |
| $A_{180}^3$ | 180 | 3 | 3.5 | 3 (4-wave) |
| $B_{180}^3$ | 180 | 3 | 7 | 3 (4-wave) |
| $C_{180}^3$ | 180 | 3 | 7 | 4 (5-wave) |

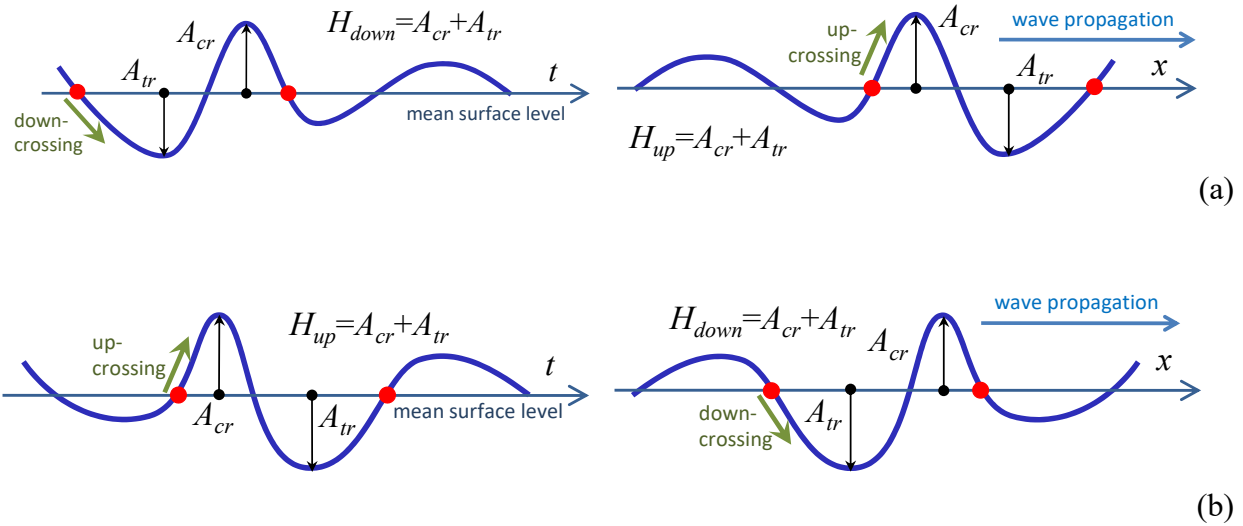

**Figure 1.** Wave configurations with the deeper trough from the frontal side, a *frontal rogue wave* (a), and from the rear side, a *rear rogue wave* (b). The sketches correspond to the zero-crossing analysis of a surface displacement record, represented by a time series (left) and by a space series (right). The individual waves are confined between the red circles denoting the zero-crossing points. The waves propagate in the direction of the coordinate axis *Ox*.

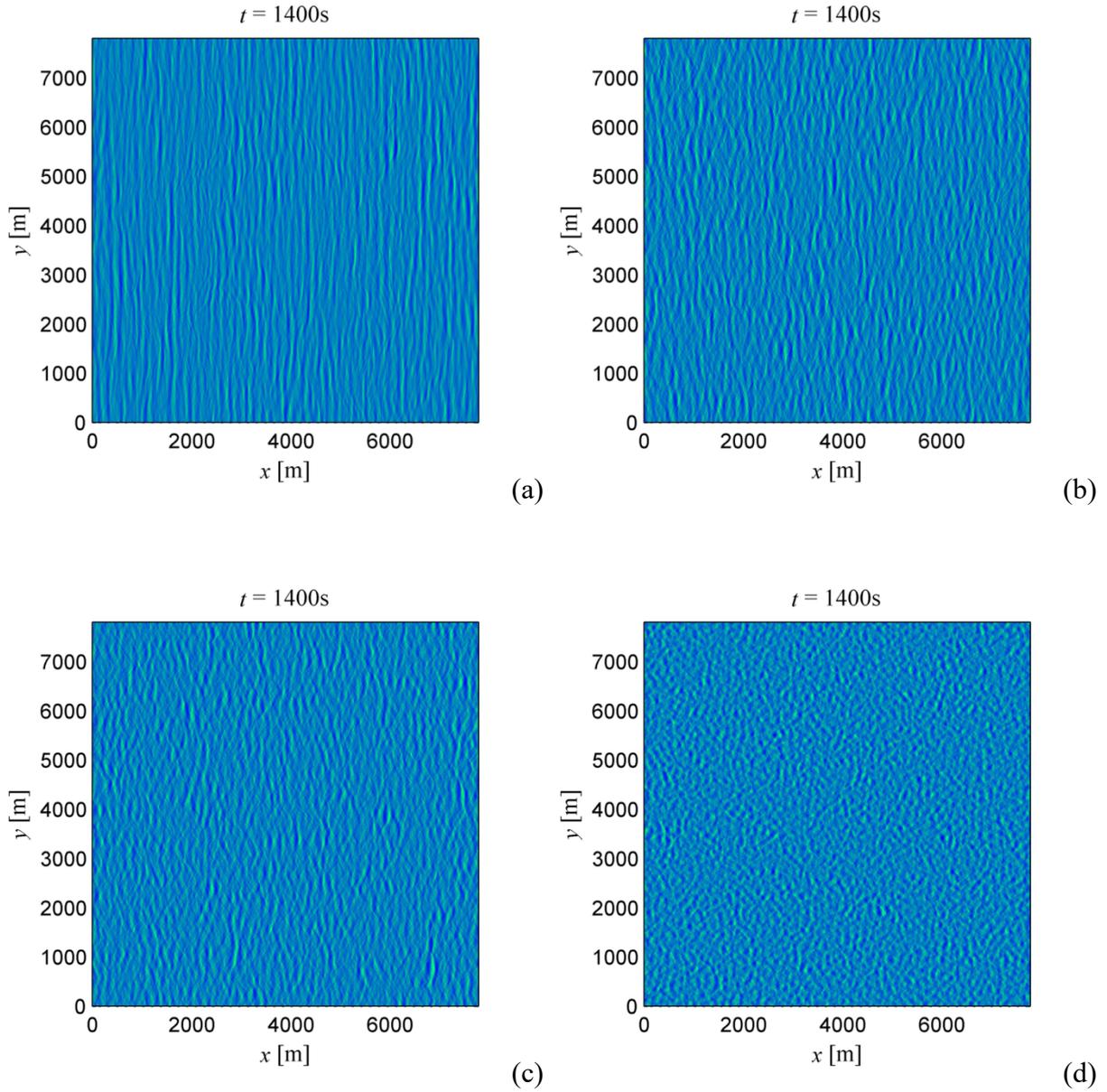

**Figure 2.** Water surfaces at the end of the nonlinear simulations, $t = 1400$ s, for the directional spread parameters $\Delta\theta = 12°$ (a), $\Delta\theta = 62°$ (b), $\Delta\theta = 90°$ (c) and $\Delta\theta = 180°$ (d). The main direction of wave propagation is rightwards.

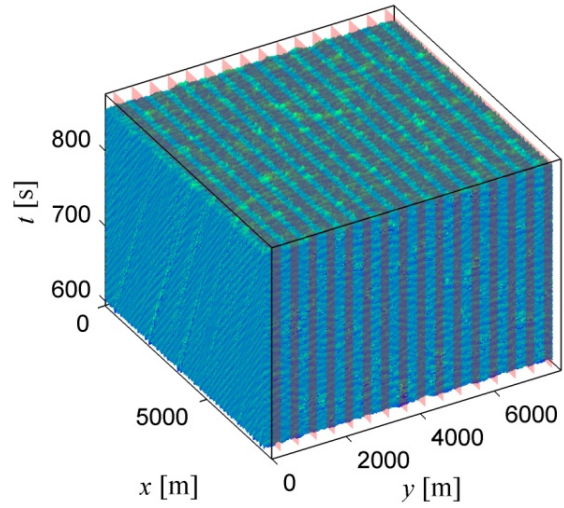

(a)

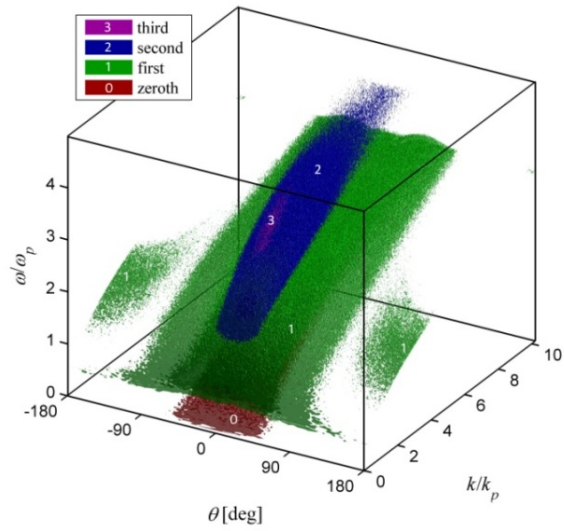

(b)

**Figure 3.** Sequence of surfaces $\eta(x,y,t)$ with cutting planes (red) showing the method of retrieving time series for the probabilistic analysis (a), and the distribution of energy in the corresponding spatio-temporal Fourier domain (b) for the experiment $B_{62}^{3}$. Isosurfaces of the Fourier amplitudes are shown; different colors denote nonlinear harmonics (see the legend for numbers $l$ of the corresponding spectral areas).

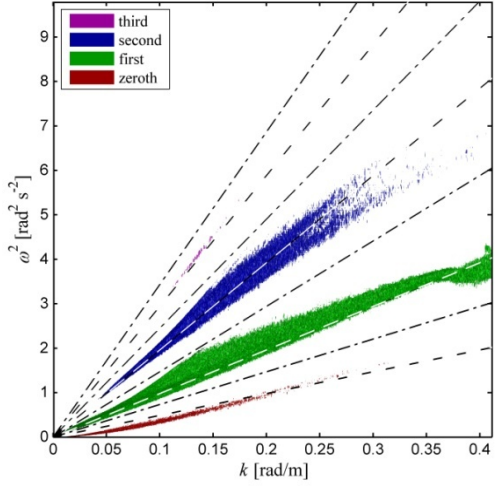 (a)
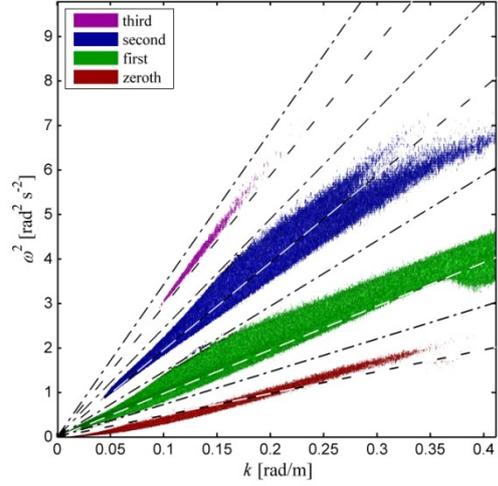 (b)
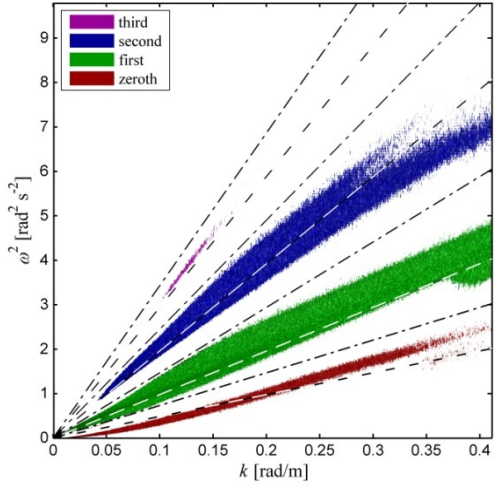 (c)
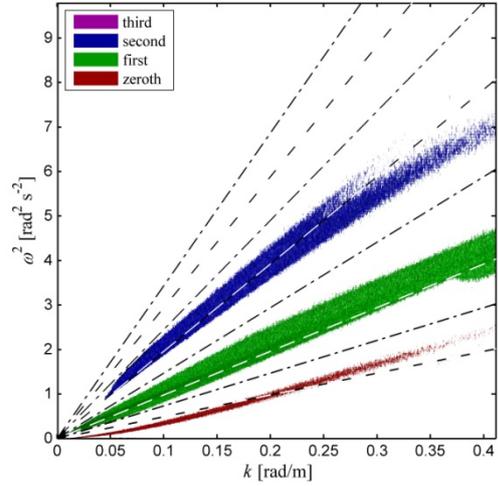 (d)

**Figure 4.** The isosurfaces similar to as in Fig. 3b, but the view from the side along the $O\theta$ axis, and the vertical axis represents the squared frequency. Results for the experiments with intense waves $C_{12}^6$ (a), $C_{62}^6$ (b), $C_{90}^3$ (d) and $C_{180}^3$ (d). The straight dashed lines correspond to the relation (6) for $n = 0.5, 1, 2, 3$ (from below upward); the straight dash-dotted lines correspond to $n = 0.75, 1.5, 2.5, 3.5$. Different colors show the separated nonlinear harmonics.

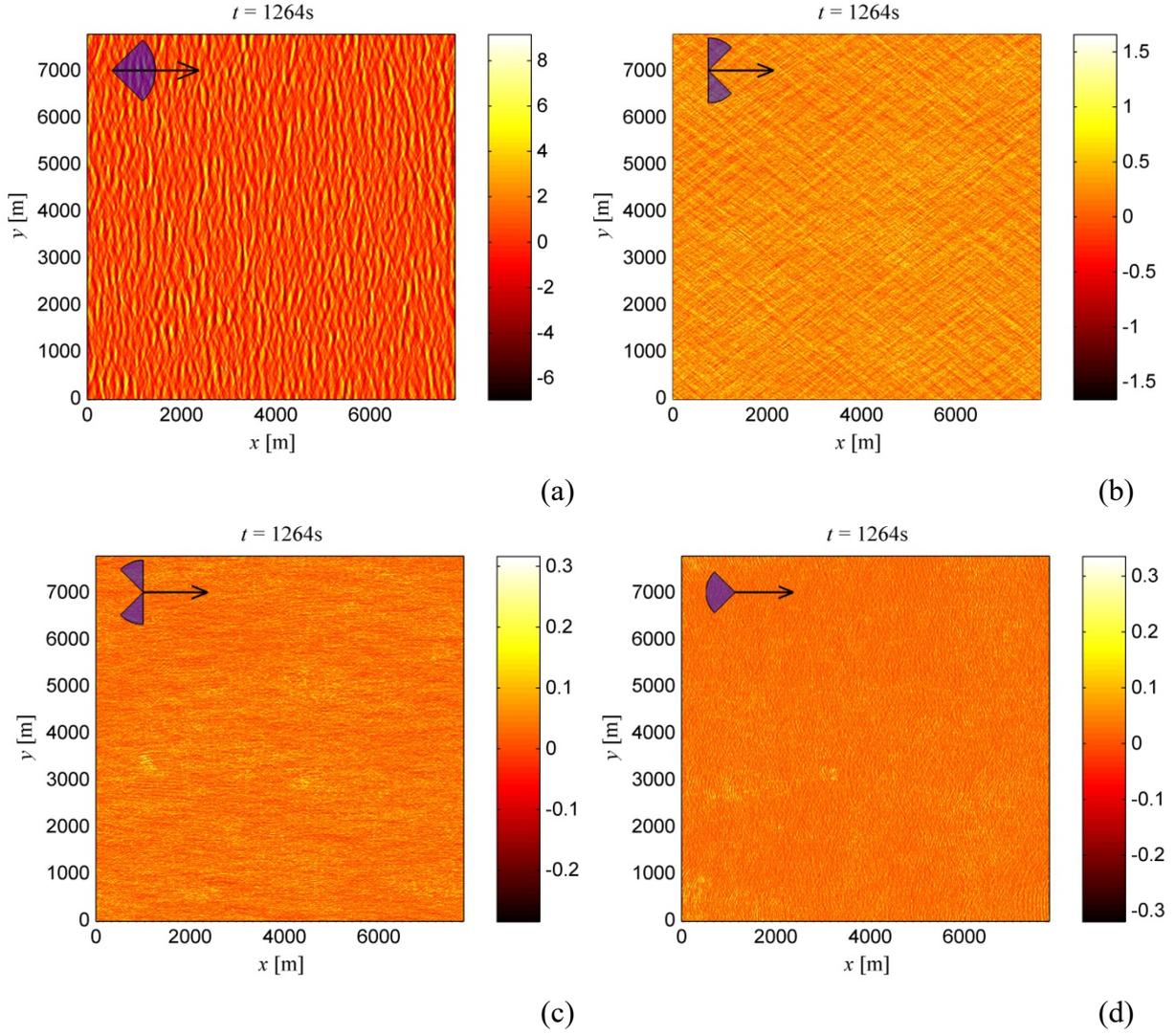

**Figure 5.** An example of decomposition of the wave surface according to four sectors of propagation angles within the intervals [0, 45º) (a), [45º, 90º) (b), [90º, 135º) (c) and [135º, 180º] (d) for the experiment $B_{62}^{3}$.

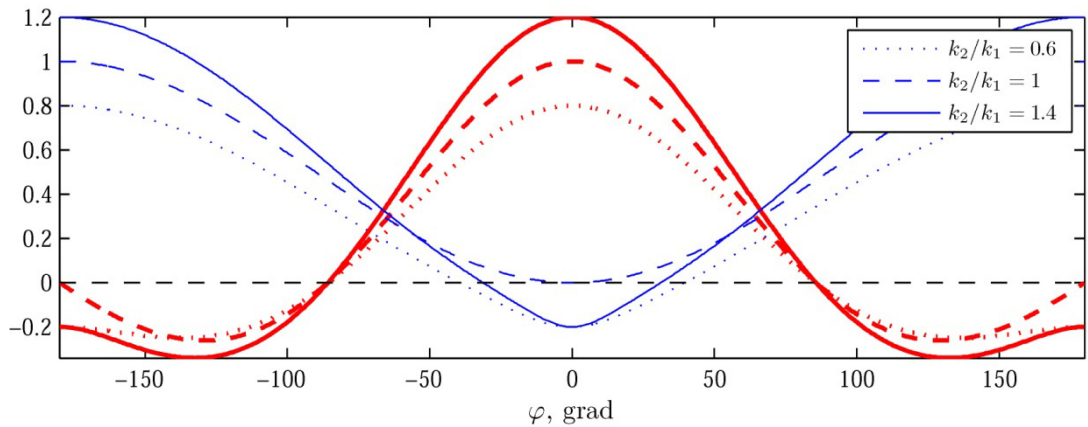

**Figure 6.** Nonlinear coefficients $B_p^{\infty}$ (red thick curves) and $B_m^{\infty}$ (blue thin curves) as functions of the angle $\varphi$ between vectors $\mathbf{k}_1$ and $\mathbf{k}_2$ for three combinations of the wave lengths $k_1$ and $k_2$ (see the legend).

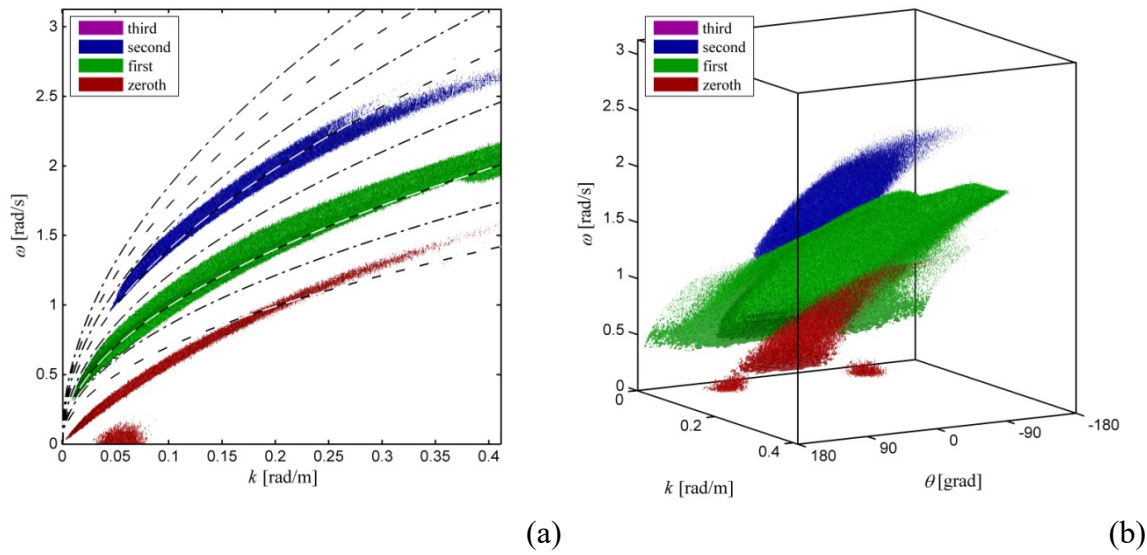

(a)                                                               (b)

**Figure 7.** Same as in Fig. 4d but with the normal frequency along the vertical axis (a), and another view of the plot (b).

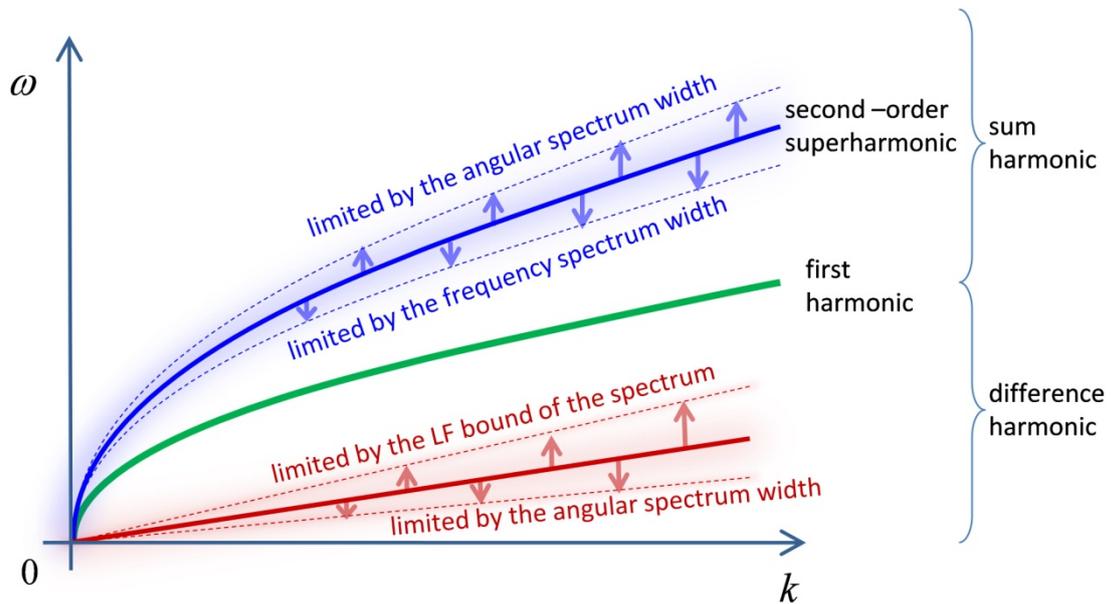

**Figure 8.** Sketch of the wave energy distribution in the ($k$, $\omega$) plane due to the quadratic nonlinearity.

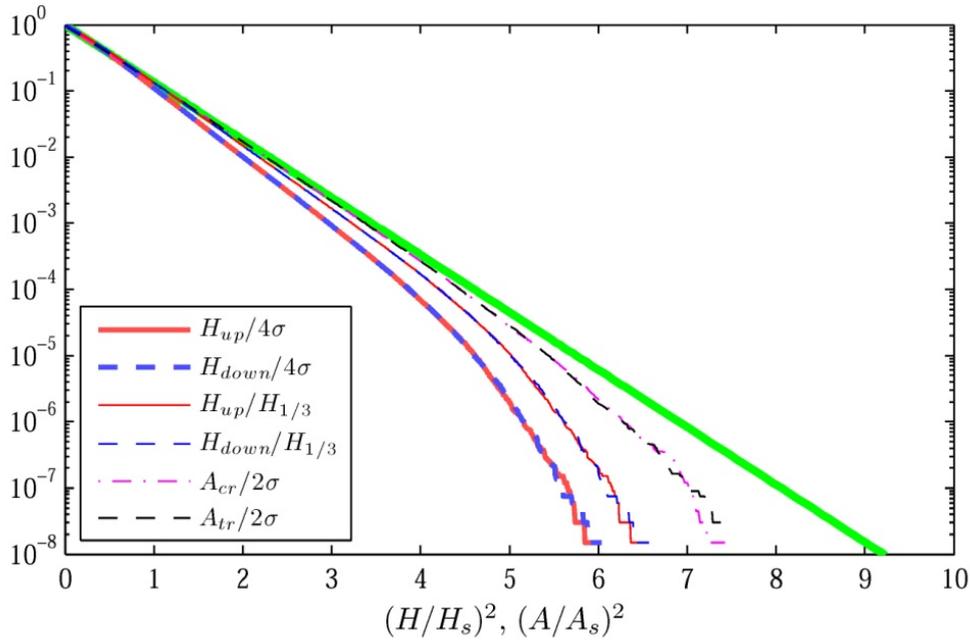

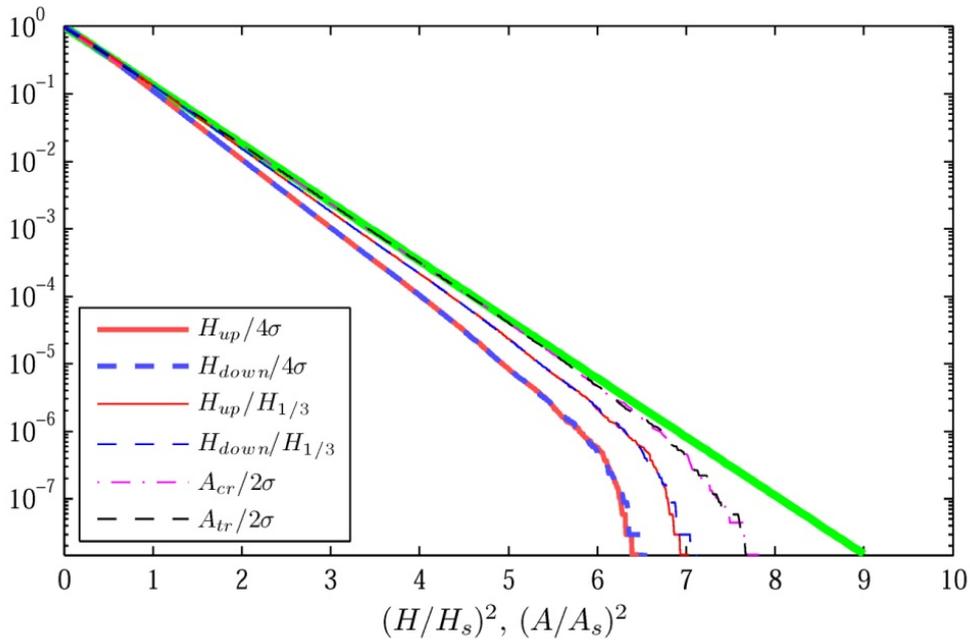

**Figure 9.** Exceedance probability distributions for waves simulated within the linear theory: long-crested waves $L_{12}^3$ (a) and short-crested waves with $L_{62}^3$ (b). The green lines correspond to the theoretical Rayleigh distribution.

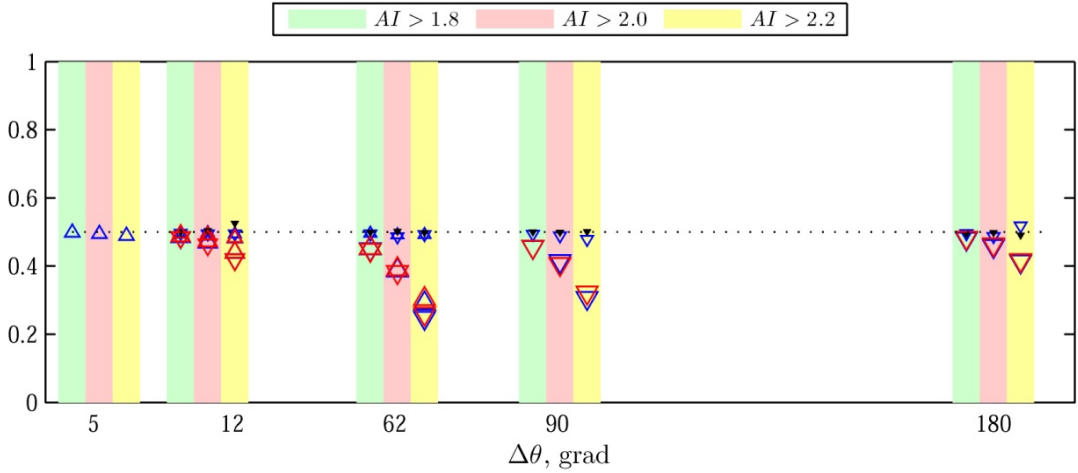

**Figure 10.** Portions of down-crossing rogue waves $P_{down}/(P_{down} + P_{up})$ in time series for three thresholds of $AI = H/(4\sigma)$. The symbol coding is as follows: the size denotes the significant wave height from 3.5 m to 7 m; the triangle orientation denotes the peakedness factor, $\gamma = 3$ ($\nabla$) and $\gamma = 6$ ($\Delta$); the symbol color denotes the nonlinearity parameter, linear simulation (black), $M = 3$ (blue) and $M = 4$ (red).

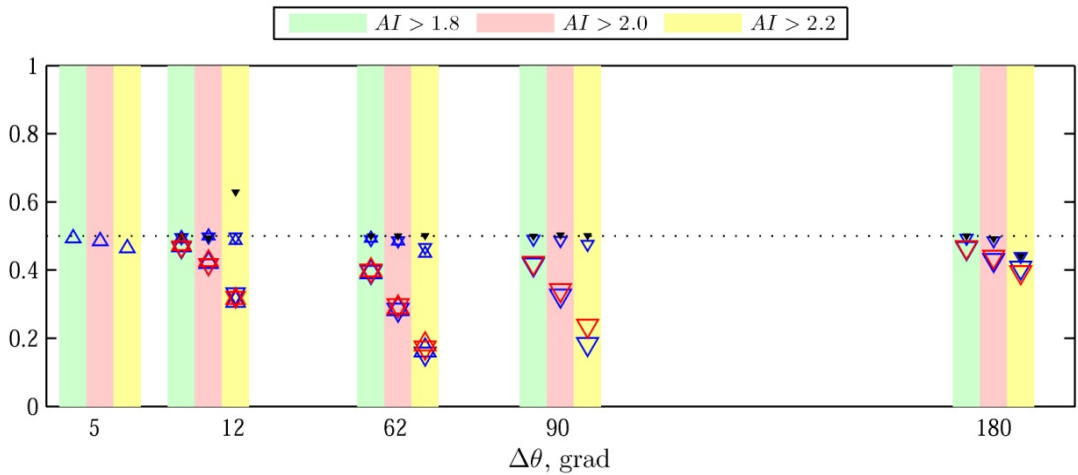

**Figure 11.** Same as in Fig. 10, but the portions of up-crossing rogue waves $P_{up}/(P_{down} + P_{up})$ in space series.

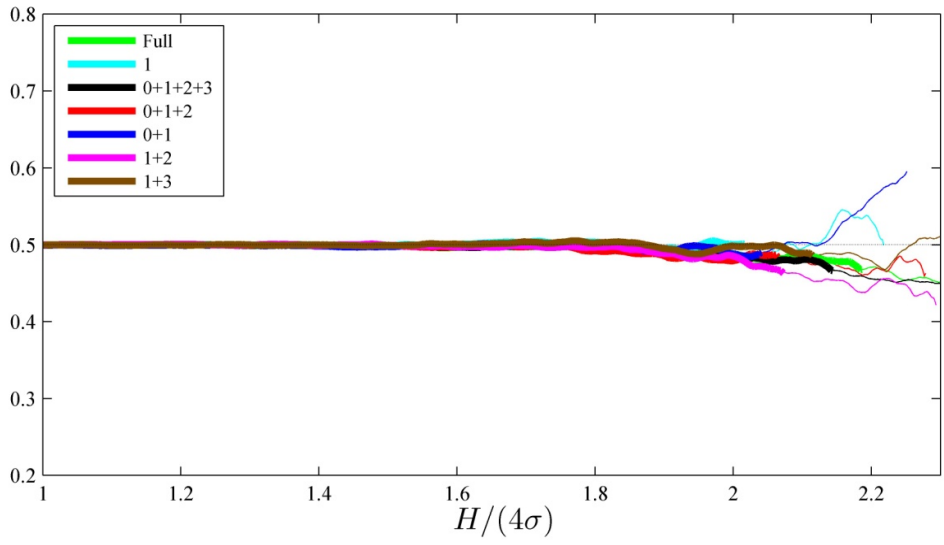

(a)

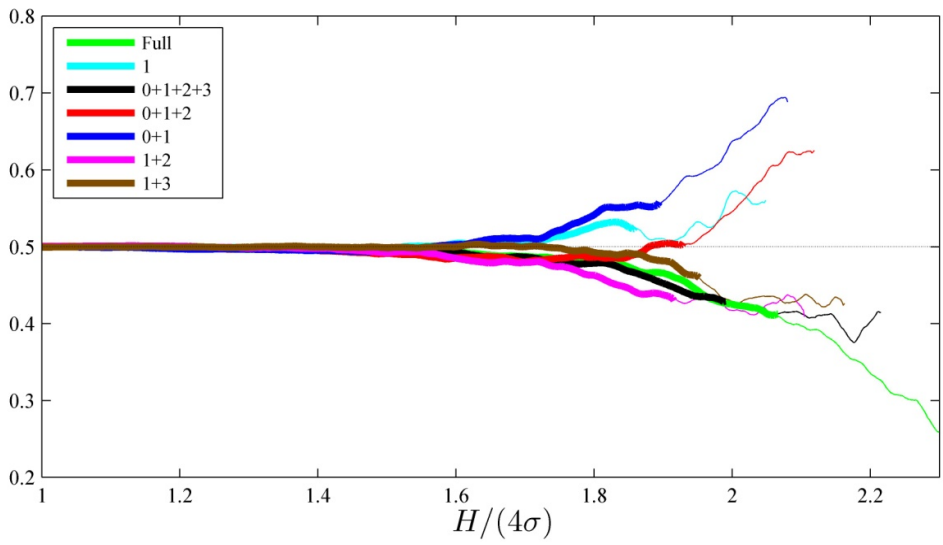

(b)

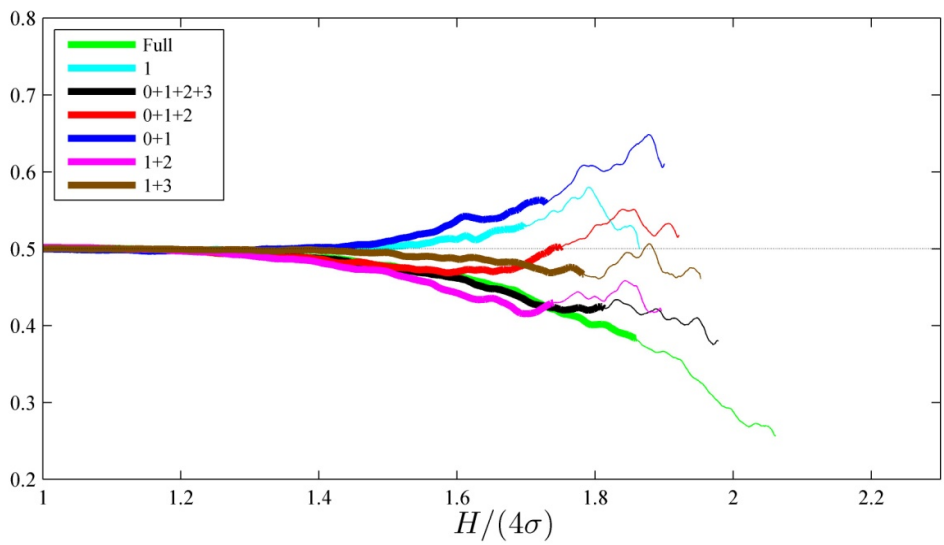

(c)

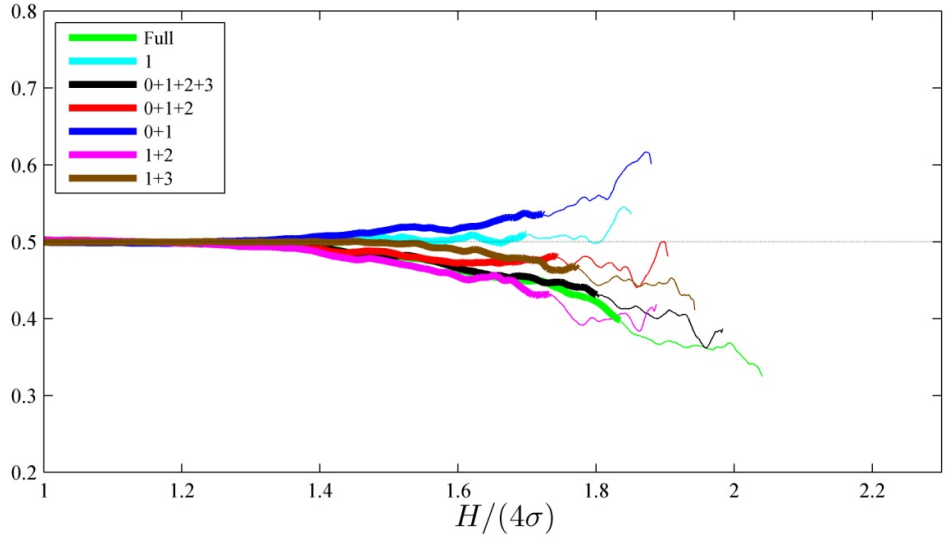

(d)

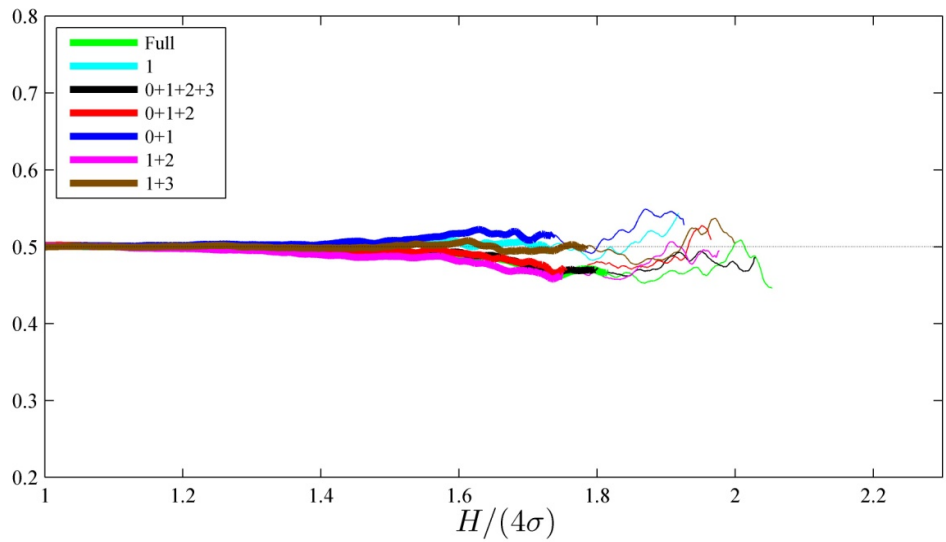

(e)

**Figure 12.** Portions of up-crossing waves $P_{up}/(P_{down} + P_{up})$ in space series as functions of the wave amplification $H/4\sigma > 1$ for different combinations of the wave harmonics (see the legend for numbers $l$). Thin lines correspond to the ensembles consisting of 50–500 individual waves, whereas thicker lines are for the ensembles of at least 500 waves. A 10-point moving average is used. The standard deviation $\sigma$ corresponds to the full wave field. Different panels show the experiments with intense waves and different directional spreads: $B_5^6$ (a), $C_{12}^6$ (b), $C_{62}^6$ (c), $C_{90}^3$ (d) and $C_{180}^3$ (e).

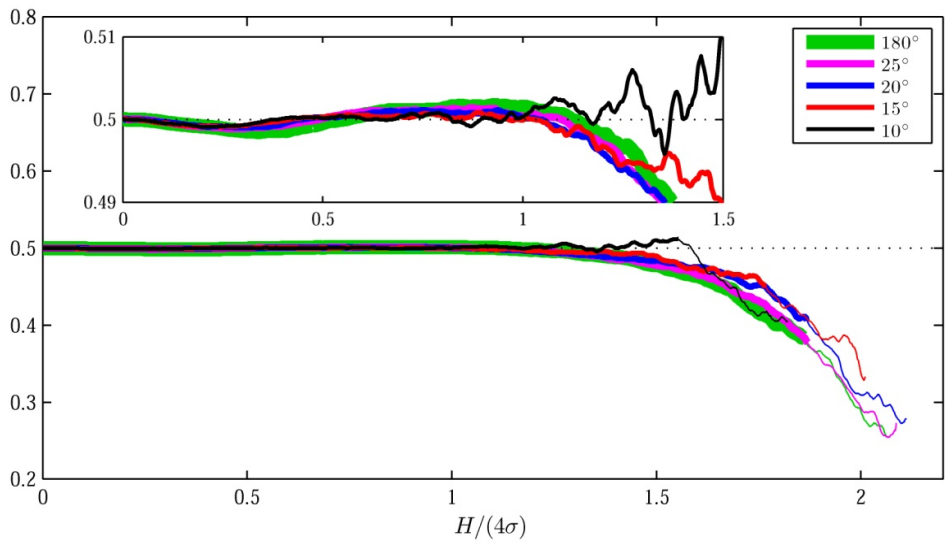

**Figure 13.** Portions of up-crossing waves $P_{up}/(P_{down} + P_{up})$ in space series of the experiment $C_{62}^6$ for waves propagating within different intervals of angles $[-\Theta, \Theta]$, see the legend for the values of $\Theta$. Thin lines correspond to the ensembles consisting of 50–500 individual waves, whereas thicker lines are for the ensembles of at least 500 waves. A 10-point moving average is used. The standard deviation $\sigma$ corresponds to the full wave field.

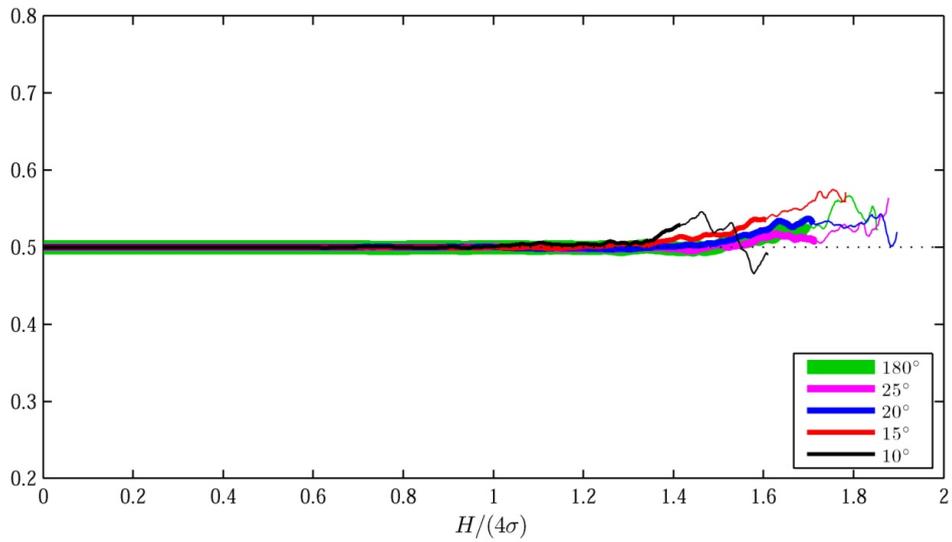

**Figure 14.** Same as in Fig. 12 but for the first harmonic component $l = 1$.